\documentclass[12pt]{article}
\begin{document}

\def\mathbi#1{\textbf{\em #1}}

\title{The Complexity of $3SAT_{N}$ and the $\mathcal P$ versus $\mathcal {NP}$ Problem}
\author{Ruijia Liao}
\date{}
\maketitle

\begin{abstract}
We introduce the $\mathcal{NP}\mbox{-}$complete problem $3SAT_{N}$
and extend Tovey's results to a classification theorem for this problem.
This theorem leads us to generalize the concept of truth 
assignments for $SAT$ to aggressive truth assignments for $3SAT_{N}$.
We introduce the concept of a set compatible with the $\mathcal P$ 
versus $\mathcal {NP}$ problem, and prove that
all aggressive truth assignments are pseudo-algorithms. We combine 
algorithm, pseudo-algorithm and diagonalization method to study the 
complexity of $3SAT_{N}$ and the $\mathcal P$ versus $\mathcal {NP}$ 
problem. The main result is $\mathcal P \neq \mathcal{NP}$.
\end{abstract}

\subsection*{1. Introduction}
In computational complexity theory, the Boolean satisfiability problem ($SAT$) 
is a decision problem, in which the instance is a Boolean expression written 
using only AND, OR, NOT, variables, and parentheses. The question is: given 
the expression, is there some assignment of $true$ and $false$ values to 
the variables that make the entire expression true? $SAT$ is the first known 
$\mathcal {NP}$-complete problem, proven by Stephen Cook in 1971 
\cite{Cook1}.  Independently in 1973, Leonid Levin showed that a variant of 
the tiling problem is $\mathcal {NP}$-complete \cite{Levin1}. In 1972, Richard 
Karp proved that several other problems were also $\mathcal {NP}$-complete 
\cite{Karp1}. In particular, the Boolean satisfiability problem remains 
$\mathcal {NP}$-complete even if all expressions are written in conjunction 
normal form with 3 variables per clause (3-CNF), yielding the $3SAT$ problem. 
We can define the $kSAT$ problem in a similar way. Let $(r,s)\mbox{-}SAT$ 
be the class of instances with exactly $r$ distinct variables per clause and 
at most $s$ occurrences per variable. In 1984 \cite{Tovey1}, Craig Tovey 
proved that every instance of $(r,r)\mbox{-}SAT$ is satisfiable and 
$(3,4)\mbox{-}SAT$ is $\mathcal {NP}$-complete.  Indeed, using these 
$(3,s)\mbox{-}SAT$ for $s = 1, \cdots, 4$ and the results in \cite{Tovey1}, 
we can easily classify all instances in $3SAT$ by polynomial time in each 
instance's length. 

In this paper, we use Tovey's idea in \cite{Tovey1} to classify the 
$\mathcal{NP}\mbox{-}$complete problem $3SAT_{N}$. We prove that 
this classification takes polynomial time. With this classification theorem, 
we introduce the concept of aggressive truth assignment. 
We also introduce the concept of a set compatible with the $\mathcal P$ 
versus $\mathcal {NP}$ problem.  Any element in a set that is compatible 
with the $\mathcal P$ versus $\mathcal {NP}$ problem is a pseudo-algorithm.
We can prove that the set of all aggressive truth assignments is 
compatible with the $\mathcal P$ versus $\mathcal {NP}$ problem,
and hence  any aggressive truth assignment is a pseudo-algorithm.
Using these pseudo-algorithms, we endow some set of algorithms with a 
metric and then introduce Cauchy sequences among them. The Cauchy 
sequences of algorithms in this paper are essentially the Cauchy sequences 
of pseudo-algorithms. Like the role of Cauchy sequences of rational 
numbers in real number theory, the Cauchy sequences of algorithms 
allow us to use pseudo-algorithms to approximate some algorithms. In 
1874,  Cantor \cite{Cantor1} established that real numbers are 
uncountable. Sixteen years later, he proved his theorem again using 
diagonal argument \cite{Cantor2}.  Surprisingly, by analyzing 
computations of some pseudo-algorithms,  we can use Cantor's 
diagonalization method to prove that there are uncountably many 
algorithms under some assumption.  It contradicts the fact that there 
are only countably many algorithms (see e.g. \cite{Hein1}).  Therefore, 
the assumption must be false.  The concept of a set compatible 
with the $\mathcal P$ versus $\mathcal {NP}$ problem is
crucial which prevents misusing the pseudo-algorithms.

In 1975,  T.Baker, J.Gill, and R.Solovary \cite{Baker1} introduced the 
following relativized worlds: there exist oracles $A$ and $B$, such that 
$\mathcal{P}^A \not= \mathcal{NP}^A$ and 
$\mathcal{P}^B = \mathcal{NP}^B$. 
They also pointed out that the relativizing method could not solve the 
$\mathcal{P}$ versus $\mathcal{NP}$ problem.  In the early 1990's, 
A.Razborov and S.Rudich \cite{Razborov1} defined a general class of 
proof techniques for circuit complexity lower bounds, called natural 
proofs. At the time all previously known circuit lower bounds were natural, 
and circuit complexity was considered a very promising approach for 
resolving $\mathcal{P} = \mathcal{NP}$. However, Razborov and Rudich 
showed that, if certain kinds of one-way functions exist, then no natural 
proof method can distinguish between $\mathcal{P}$ and $\mathcal{NP}$. 
Although one-way functions have never been formally proven to exist, 
most mathematicians believe that they do.  Thus it is unlikely that natural 
proofs alone can resolve $\mathcal{P} = \mathcal{NP}$. In 1992,  
A.Shamir \cite{Shamir1} used a new non-relativizing technique to prove 
$IP = PSPACE$.  However, in 2008, S.Aaronson and A.Wigderson 
\cite{Aaronson1} showed that the main technical tool used in the 
$IP = PSPACE$ proof, known as arithmetization, was also insufficient 
to resolve $\mathcal{P} = \mathcal{NP}$. In this paper, each 
pseudo-algorithm can be explicitly expressed by Algorithm 1 or Algorithm 2,
and Algorithm 3; however, it is not an algorithm.  It takes finite steps to 
partially evaluate any $\eta \in 3SAT$ and, most importantly, 
it is different from any oracle and arithmetization.  Since the new 
argument combines algorithm,  pseudo-algorithm and diagonalization 
method, it circumvents relativization, natural proofs and algebrization.

In Section 2, we give some definitions and notations, we describe two similar 
algorithms and show that one is always more efficient than another. Using
this result and the same argument, we prove in Section 8 that any element 
in some algorithm sequence is always more efficient than some polynomial 
time algorithm.  We prove that $3SAT_{N}$ is an 
$\mathcal{NP}\mbox{-}$complete problem in Section 3. We extend Tovey's 
results in \cite{Tovey1} to a classification theorem in $3SAT_{N}$ and 
define another algorithm in Section 4. We generalize the concept of truth 
assignment to aggressive truth assignments in Section 
5 based on the Classification Theorem. We define the composition of two 
or more aggressive truth assignments and investigate, 
$\mathcal{TA}^{\infty}$, a set of all aggressive truth assignments under 
this operation. We introduce the concept of distance between any two 
elements in $\mathcal{TA}^{\infty}$  and endow it with a metric.

In Section 6, we extend this metric concept to $< f >$, a set of 
algorithms generated by algorithm $f$ and aggressive truth assignments, 
in which we define Cauchy sequences. We give the definition of 
pseudo-algorithms.  In Section 7, we discuss equivalence of algorithms 
and pseudo-algorithms and set up an equivalence relation in $< f >$.
Any two elements in $\mathcal{TA}^1$ are always equivalent.  However, 
$\mathcal{TA}^2$ has infinitely many equivalence classes.  In Section 8, 
we introduce the concept of regular Cauchy sequence in $< f > ^2$.  
It is critically important to be able to construct an algorithm $f_\zeta$ 
from $\{ f_n \}$, an arbitrarily given regular Cauchy sequence. Such 
$f_\zeta$ is called the representation of $\{ f_n\}$. We generalize the
concept of equivalence relation from $< f >$ to the set of all regular 
Cauchy sequences in $< f > ^2$, and prove that 
any two regular Cauchy sequences in different equivalence classes have 
different representations and represent two different algorithms. For 
any regular Cauchy sequence $\{ f a_n a_0 \}$,  there exists a 
polynomial time algorithm $f \tilde{a_0} \tilde{a_0}$ which is less efficient 
than $f a_n a_0$ for $n = 1, 2, \cdots$, thus, its representation 
$f_\zeta$ is always a polynomial time algorithm.  We show that the 
set of all aggressive truth assignments is a compatible set with the 
$\mathcal P$ versus $\mathcal {NP}$ problem in Section 9.  Using the 
regular Cauchy sequences and diagonalization method,  we prove that 
$\mathcal{P} \not= \mathcal{NP}$ in Section 10.

\subsection*{2. Preliminary}
Let $SAT(n)(x_1, \cdots, x_n)$ be the set of all expressions in $SAT$ in
which each element uses exactly n variables and their negations 
$\{x_1, \cdots, x_n, \neg x_1, \cdots, \neg x_n\}$.
From the definition, $SAT(n) \cap SAT(n+1) = \emptyset$ and
$\cup^{\infty}_{n=3} SAT(n) \subseteq SAT$. For $r \le n$ and
$1 \le s$, let $(r,s)\mbox{-}SAT(n) = (r,s)\mbox{-}SAT \cap SAT(n)$.
We can show that, for any $\eta \in SAT$, there exists a polynomial time
algorithm in the length of $\eta$ to find the integer
$n$ such that $\eta$ is generated by variables and their negations 
$\{x_{i_1}, \cdots, x_{i_n}, \neg x_{i_1}, \cdots, \neg x_{i_n}\}$ where
$i_1 < \cdots < i_n$. Furthermore, there exists a linear time map 
$\phi_{map}$, such that 
$\phi_{map}(x_{i_k}) = x_k, \phi_{map}(\neg x_{i_k}) = \neg x_k$,
for $k = 1, \cdots, n$ and $\tilde\eta = \phi_{map}(\eta) \in SAT(n)$.
Thus, if $\eta \in (r,s)\mbox{-}SAT$, then
$\tilde\eta = \phi_{map}(\eta) \in (r,s)\mbox{-}SAT(n)$ for some $n$.
Obviously, $\tilde\eta$ is satisfiable if and only if $\eta$ is
satisfiable, and the numbers of clauses of $\tilde\eta$ and $\eta$ are
the same. For any $\eta \in SAT(n)$, the above algorithm is trivial 
and the map $\phi_{map}$ can be viewed as the identical map. In the 
following discussion, for convenience, we use $\phi_{map}$
to represent the above algorithm and the map.

The map $x_i^{\ast}$ from $\{x_j, \neg x_j\}$ to \{{\it true, false, undef}\} is
defined as $x_i^{\ast}(x_i) = true, x_i^{\ast}(\neg x_i) = \mbox{{\it false}}$,
$x_i^{\ast}(x_j) = \mbox{{\it undef}}$ and $x_i^{\ast}(\neg x_j) = \mbox{{\it undef}}$
 if $i \not= j$. The map $\neg x_i^{\ast}$ is defined in a similar way.
An atomic truth assignment $e_i$ is defined as $e_i = x_i^{\ast}$ or  
$e_i = \neg x_i^{\ast}$. The negation operator can be applied to $e_i$
too, i.e., if $e_i = x_i^{\ast}$, $\neg e_i = \neg x_i^{\ast}$, and if 
$e_i = \neg x_i^{\ast}$, $\neg e_i = x_i^{\ast}$. 

Let 
\begin{eqnarray}
\eta = (y_{1,1} \vee \cdots \vee y_{1,j_1}) \wedge 
(y_{2,1} \vee \cdots \vee y_{2,j_2}) \wedge \cdot \cdot \cdot \wedge
(y_{m,1} \vee \cdots \vee y_{m,j_m}) \in SAT(n),
\label{eqn1.1}
\end{eqnarray}
where $y_{i,j} = x_k \mbox{ or } \neg x_k$ for some $k$, $\mbox{ the subscript }\{i,j,k\} \in \{1, 2, \cdots, n\}$.
A truth assignment $e_1 e_2 \ldots e_n$ is defined as
\begin{eqnarray}
e_1 e_2 \ldots e_n(\eta)
 &=& (e_{1,1}(y_{1,1}) 
  \vee \cdots \vee e_{1,j_1}(y_{1,j_1})) \wedge \nonumber \\
&&(e_{2,1}(y_{2,1}) 
  \vee \cdots \vee e_{2,j_2}(y_{2,j_2})) \wedge \cdots \wedge \nonumber \\
&&(e_{m,1}(y_{m,1}) 
  \vee \cdots \vee e_{m,j_m}(y_{m,j_m}))
 \in \{true, false\}. \label{eqn1.2}
\end{eqnarray}
We can use this definition to describe an efficient algorithm for the 
truth assignment $e_1 e_2 \ldots e_n$ to evaluate any $\eta \in SAT(n)$. 

\noindent $\bf Algorithm\mbox{ }1$.\\ 
\indent for $k = 1, 2, \cdots, m$ \\
\indent\indent evaluate clause $(y_{k,1} \vee \cdots \vee y_{k,j_k})$ as follows:\\ 
\indent\indent for $l = 1, 2, \cdots, j_k$ \\
\indent\indent\indent for $p = 1, 2, \cdots, n$ \\
\indent\indent\indent\indent if $e_p(y_{k,l}) = \mbox{\it{undef}}$, continue to the inner loop, 
i.e., try next $e_{p+1}$; \\
\indent\indent\indent\indent if $e_p(y_{k,l}) = \mbox{{\it false}}$ and $l < j_k$, 
continue to the middle loop, i.e., try next $y_{k,l+1}$;\\
\indent\indent\indent\indent if $e_p(y_{k,l}) = \mbox{{\it false}}$ and $l = j_k$, return {\it false};\\
\indent\indent\indent\indent if $e_p(y_{k,l}) = true$\\
\indent\indent\indent\indent\indent if $k < m$, continue to the outer most loop, i.e., try next clause;\\
\indent\indent\indent\indent\indent if $k = m$, return $true$.

The elementary steps of Algorithm 1 are the evaluations $e_i(y_j)$, 
for $i, j = 1, \cdots, n$, and returning {\it true} or {\it false}. It is a
polynomial time algorithm. In this paper,  for simplicity, we do not count 
steps of comparison among $e_i(y_j)$ and {\it true, false, undef}, 
and loop indexes and variables' subscripts. The following algorithm is 
similar to Algorithm 1 but less efficient.

\noindent $\bf Algorithm\mbox{ }2$.\\ 
\indent set $c_g = 0$;\\
\indent for $k = 1, 2, \cdots, m$ \\
\indent\indent evaluate clause $(y_{k,1} \vee \cdots \vee y_{k,j_k})$ as follows:\\ 
\indent\indent $c_l = 0$;\\
\indent\indent for $l = 1, 2, \cdots, j_k$ \\
\indent\indent\indent for $p = 1, 2, \cdots, n$ \\
\indent\indent\indent\indent if $e_p(y_{k,l}) = \mbox{\it{undef}}$, continue to the inner loop, 
i.e., try next $e_{p+1}$; \\
\indent\indent\indent\indent if $e_p(y_{k,l}) = \mbox{{\it false}}$ and $l < j_k$, 
continue to the middle loop, i.e., try next $y_{k,l+1}$;\\
\indent\indent\indent\indent if $e_p(y_{k,l}) = \mbox{{\it false}}$ and $l = j_k$\\
\indent\indent\indent\indent\indent if $c_l = 0$, increase $c_g$ by 1;\\
\indent\indent\indent\indent\indent if $k < m$, continue to the outer most loop, i.e., try next clause;\\
\indent\indent\indent\indent\indent if $k = m$, return $c_g$;\\
\indent\indent\indent\indent if $e_p(y_{k,l}) = true$,  increase $c_l$ by 1\\
\indent\indent\indent\indent\indent if $l = j_k$ and $k < m$, continue to the outer most loop, i.e., try next clause;\\
\indent\indent\indent\indent\indent if $l = j_k$ and $k = m$, return $c_g$.

The elementary steps of Algorithm 2 are the evaluations $e_i(y_j)$, 
for $i, j = 1, \cdots, n$, setting $c_\alpha$ to 0 and increasing 
$c_\alpha$ by 1 for $\alpha = g, l$,  and returning $c_g$. It is also 
a polynomial time algorithm.  The return value $c_g$ equals $0$ if 
and only if the algorithm evaluates $\eta$ as true.

The significant different steps between Algorithm 1 and Algorithm 2 
are returning {\it true}, returning {\it false}, returning $c_g$, setting 
$c_\alpha = 0$ and increasing $c_\alpha$ by 1 for $\alpha = g, l$.  
To complete evaluate a clause,  Algorithm 2 must inspect each variable 
in the clause for all cases while Algorithm 1 needs 
get only one variable with true evaluation value for the case of evaluated true. 
To complete evaluate a whole expression,  Algorithm 2 must evaluate 
each clause while Algorithm 1 needs getting only one clause with false 
evaluation value for the case of evaluated false.  In other words, 
Algorithm 2 always takes more steps than Algorithm 1 does.  Thus, 
we have the following:

\noindent $\bf Lemma\mbox{ }1$. For any given two truth assignments 
$e^1_1 e^1_2 \ldots e^1_n$ and $e^2_1 e^2_2 \ldots e^2_n$, and
any instance $\eta \in SAT(n)$. Let $s_1$ and $s_2$ be the number of 
steps of $e^1_1 e^1_2 \ldots e^1_n$ and $e^2_1 e^2_2 \ldots e^2_n$ 
using Algorithm 1 and Algorithm 2 to evaluate $\eta$ respectively, then 
$s_1 < s_2$.

From the definition, a truth assignment $e_1e_2 \cdots e_n$
is determined by each atomic truth assignment $e_i$, it is 
independent of their order. However, for convenience,  it 
always takes ascending order.  For any integers $m \ge n \ge 3$, 
we can apply the truth assignment $e_1e_2 \cdots e_m$ to any 
instance of $SAT(n)$. If $m > n$, we use $e_1, e_2, \cdots, e_n$ 
and ignore $e_{n+1}, \cdots, e_m$. Similarly, we can define a 
generalized truth assignment $e_1e_2 \cdots e_n \cdots$ and 
apply it to any instance of $SAT(n)$ for any integer $n \ge 3$. 
However, the generalized expression is defined by finite 
information. For example, for any given integer $k > 0$,
\begin{eqnarray}
e_1e_2 \cdots e_k \cdots &=&
e_1e_2 \cdots e_kx^{\ast}_{k+1}x^{\ast}_{k+2}x^{\ast}_{k+3}x^{\ast}_{k+4} \cdots, \mbox{ or} \nonumber \\
&&e_1e_2 \cdots e_kx^{\ast}_{k+1}\neg x^{\ast}_{k+2}x^{\ast}_{k+3}\neg x^{\ast}_{k+4} \cdots, \mbox{ or} \nonumber \\
&&e_1e_2 \cdots e_k\neg x^{\ast}_{k+1}\neg x^{\ast}_{k+2}\neg x^{\ast}_{k+3}\neg x^{\ast}_{k+4}\cdots. \nonumber
\end{eqnarray}
The first expression is called the positive extension of $e_1e_2 \cdots e_k$ 
and the last one is called the negative extension of $e_1e_2 \cdots e_k$.  
A generalized truth assignment $e_1 e_2 \cdots e_n \cdots$ is called 
negative if $e_n = \neg x_n^{\ast}$ for any integer $n \ge 1$.

\noindent $\bf Remark\mbox{ }1.$  Define 
$\Phi(e_1e_2 \cdots e_k \cdots) = 0.\phi(e_1) \phi(e_2) \cdots \phi(e_k) \cdots$,
where $\phi(x_i^{\ast}) = 1$ and $\phi(\neg x_i^{\ast})$ $ = 0$ for 
$i = 1, 2, \cdots$.  The assumption of  finite information means that $\Phi$ 
maps a generalized truth assignment to a binary number in $[0, 1]$ which has 
finite many digit ones or infinite many digit ones but cyclic after some digits. 
In other words, $\Phi$ maps any generalized truth assignment to a rational 
number in $[0, 1]$. In the worst case, $\Phi$ maps two different generalized 
truth assignments to one binary number.  Thus, there exist only countably 
many generalized truth assignments.  In particular, for any truth assignment 
$e_1 e_2 \cdots e_n$, there are countably many generalized truth 
assignments associated with it.

\noindent $\bf Example\mbox{ }1$. Let $e_1 = x_1^{\ast}, e_2 = \neg x_2^{\ast}, e_3 = \neg x_3^{\ast}$,
$e_4 = x_4^{\ast}$,
$\eta_1 = (\neg x_1 \vee \neg x_2 \vee \neg x_3) \wedge (\neg x_1 \vee x_2 \vee x_3)$,
and $\eta_2 = (x_1 \vee x_3) \wedge (\neg x_1 \vee \neg x_3)$, then
\begin{eqnarray}
e_1e_2e_3e_4(\eta_1) &=& (e_1(\neg x_1) \vee e_2(\neg x_2) \vee e_3(\neg x_3))
\wedge (e_1(\neg x_1) \vee e_2(x_2) \vee e_3(x_3)) \nonumber \\
&=& (false \vee true \vee  e_3(\neg x_3)) \wedge  (e_1(\neg x_1) \vee e_2(x_2) \vee e_3(x_3)) \nonumber \\
&=& (e_1(\neg x_1) \vee e_2(x_2) \vee e_3(x_3) ) \nonumber \\
&=& (false \vee false \vee false) \nonumber \\ 
&=& false, \nonumber
\end{eqnarray}
and
\begin{eqnarray}
e_1e_2e_3e_4(\eta_2) &=& (e_1(x_1) \vee e_3(x_3))
\wedge (e_1(\neg x_1) \vee e_3(\neg x_3)) \nonumber \\ 
&=& (true \vee e_3(x_3)) \wedge (e_1(\neg x_1) \vee e_3(\neg x_3)) \nonumber \\ 
&=& (e_1(\neg x_1) \vee e_3(\neg x_3)) \nonumber \\
&=& (false \vee true) \nonumber \\
&=& true. \nonumber
\end{eqnarray}

\subsection*{3. An $\mathcal {NP}$-complete Problem}
Let $\eta$ be an instance of $kSAT$. A clause
$y_{i_1} \vee y_{i_2} \vee \cdots \vee y_{i_k}$ of $\eta$  
is called a tautological clause if there are 
$i_\alpha$, $i_\beta$ and variable $x$ such that 
$y_{i_\alpha} = x$ and $y_{i_\beta} = \neg x$.
A clause $y_{i_1} \vee y_{i_2} \vee \cdots \vee y_{i_k}$ 
of $\eta$ is called a full clause if it has $k$ distinct 
variables or their negations. In other words, a full clause 
has no repeated variable. An expression is called a normal 
expression if it satisfies the following conditions:
(1) it has no tautological clause,
(2) it has no repeated clause and
(3) each clause is full. 
Let 
\begin{eqnarray}
&&kSAT_N = \{\eta \mbox{ }|\mbox{ }\eta \in kSAT \mbox{ and } 
\eta \mbox{ is a normal expression}\}. \label{eqn3.1}
\end{eqnarray}
We have the following:

\noindent $\bf Theorem\mbox{ }1$. $3SAT_N$ is $\mathcal {NP}$-complete.

\noindent $\it Proof$. 
First, we show that $3SAT_N$ is in $\mathcal{NP}$. A 
nondeterministic polynomial time Turing machine can guess
a truth assignment to a given expression $\eta \in 3SAT_N$
and accept if the assignment satisfies $\eta$.  Next, we want to
prove that any given $\eta \in 3SAT$ can be reduced to 
$\tilde \eta \in 3SAT_N$ in  polynomial time in the length of 
$\eta$.  For any $\eta \in 3SAT$,

(1) If $\eta = \tilde \eta \wedge \theta$, where $\theta$ 
is a tautological clause, then $\tilde \eta$ is satisfiable 
if and only if $\eta$ is satisfiable.  Remove all 
tautological clauses. Let $\eta_1$ be the result expression.

(2) If $\eta_1 = \tilde \eta_1 \wedge \theta \wedge \theta$, 
where $\theta$ is a repeated clause. Let 
$\overline\eta_1 = \tilde\eta_1 \wedge \theta$,
then $\overline \eta_1$ is satisfiable if and only if 
$\eta_1$ is satisfiable. If $\overline \eta_1$ has any 
repeated clause, repeat this subroutine for 
$\eta_1 = \overline \eta_1$.  Let $\eta_2$ be the result 
expression.

(3) If $\eta_2$ has a clause with repeated variable, remove 
one or two these repeated variables to get a new clause 
without any repeated variable, repeat this subroutine for all 
clauses with repeated variable.  Let $\eta_3$ be the result 
expression.

(4) If $\eta_3 = \tilde \eta_3 \wedge \theta$, where 
clause $\theta = x$ or $\theta = x \vee y$. If $\theta = x$, 
we can force $x$ to be true by means of the clauses below:
\begin{eqnarray}
&&(x \vee a^{(x)}_{i,j} \vee b^{(x)}_{i,j}), \mbox{ $j = 1, 2$,} \nonumber \\
&&(d^{(x)}_{i,j} \vee \neg a^{(x)}_{i,j} \vee \neg b^{(x)}_{i,j}),
(d^{(x)}_{i,j} \vee \neg a^{(x)}_{i,j} \vee b^{(x)}_{i,j}), 
(d^{(x)}_{i,j} \vee a^{(x)}_{i,j} \vee \neg b^{(x)}_{i,j}), \mbox{ $j = 1, 2$,} \nonumber \\
&&(\neg d^{(x)}_{i,1} \vee \neg d^{(x)}_{i,2} \vee \neg x). \nonumber 
\end{eqnarray}
In order to make the above nine clauses satisfiable, 
$x$ must be true. Suppose that $x = false$, then 
$\neg d^{(x)}_{i,1} = true$ or
$\neg d^{(x)}_{i,2} = true$ by the last clause above.
If $\neg d^{(x)}_{i,j} = true$, then the trueness of
all three clauses
$(d^{(x)}_{i,j} \vee \neg a^{(x)}_{i,j} \vee \neg b^{(x)}_{i,j})$,
$(d^{(x)}_{i,j} \vee \neg a^{(x)}_{i,j} \vee b^{(x)}_{i,j})$ and 
$(d^{(x)}_{i,j} \vee a^{(x)}_{i,j} \vee \neg b^{(x)}_{i,j})$ implies that
$a^{(x)}_{i,j} \vee b^{(x)}_{i,j} = false$, and if
$(x \vee a^{(x)}_{i,j} \vee b^{(x)}_{i,j}) = true$,
we must have $x = true$. This contradicts our 
assumption. Let $\kappa$ be the wedge of the above 
nine clauses. Let 
$\overline\eta_3 = \tilde \eta_3 \wedge \kappa$,
i.e., replace clause $\theta$ with the wedge of the 
above nine clauses. Then $\overline\eta_3$ is 
satisfiable if and only if $\eta_3$ is satisfiable.

If $\theta = x \vee y$, we can force $x \vee y$ to 
be true by means of the clauses below:
\begin{eqnarray}
&&(x \vee y \vee a^{(x,y)}), \nonumber \\
&&(d^{(x,y)} \vee \neg a^{(x,y)} \vee \neg b^{(x,y)}), 
(d^{(x,y)} \vee \neg a^{(x,y)} \vee b^{(x,y)}), 
(d^{(x,y)} \vee a^{(x,y)} \vee \neg b^{(x,y)}),  \nonumber \\
&&(\neg d^{(x,y)} \vee x \vee y). \nonumber 
\end{eqnarray}
In order to make the above five clauses satisfiable, 
$x \vee y$ must be true. Suppose that $x \vee y = false$, 
then $\neg d^{(x,y)} = true$ by the last clause above. 
If $\neg d^{(x,y)} = true$, then the trueness of all 
three clauses
$(d^{(x,y)} \vee \neg a^{(x,y)} \vee \neg b^{(x,y)})$,
$(d^{(x,y)} \vee \neg a^{(x,y)} \vee b^{(x,y)})$ and 
$(d^{(x,y)} \vee a^{(x,y)} \vee \neg b^{(x,y)})$ implies that
$a^{(x,y)} \vee b^{(x,y)} = false$, and if
$(x \vee y \vee a^{(x,y)}) = true$,
we must have $x \vee y = true$. This contradicts our 
assumption. Let $\kappa$ be the wedge of the above five 
clauses. Let
$\overline\eta_3 = \tilde \eta_3 \wedge \kappa$,
i.e., replace clause $\theta$ with the wedge of the above 
five clauses. Then $\overline\eta_3$ is satisfiable if 
and only if $\eta_3$ is satisfiable. If $\overline\eta_3$ 
has any non-full clause, repeat this subroutine for 
$\eta_3 = \overline\eta_3$.  Let $\eta_4$ be the result 
expression.

Now expression $\eta_4$ is a normal expression and 
$\eta_4$ is satisfiable if and only if $\eta$ is 
satisfiable. Clearly, all subroutines (1), (2), (3) 
and (4) together take polynomial time in the length of 
$\eta$. This completes the proof of  Theorem 1.

For $n \ge 3, n \ge k, n \ge r$ and $s \ge 1$, define
\begin{eqnarray}
&&kSAT_{N}(n) = kSAT(n) \cap kSAT_{N}, \nonumber \\
&&SAT_{N}(n) = SAT(n) \cap SAT_{N}, \label{eqn3.2} \\
&&(r,s)\mbox{-}SAT_{N}(n) = (r,s)\mbox{-}SAT(n) \cap SAT_{N}(n).\nonumber
\end{eqnarray}
We can define $(r,s)\mbox{-}SAT_{N}$ in a similar way. We prove that 
$(3,4)\mbox{-}SAT_{N}$ is $\mathcal {NP}$-complete in next section.

\subsection*{4. A Classification Theorem}

In this section, we prove the following classification theorem:

\noindent $\bf Theorem\mbox{ }2$.
For every instance $\eta$ of $3SAT_{N}$,
one of the following statements is true:\\
\mbox{ }(1) $\eta \in (3,1)\mbox{-}SAT_{N}$ and $\eta$ is satisfiable,\\
\mbox{ }(2) $\eta \in (3,2)\mbox{-}SAT_{N}$ and $\eta$ is satisfiable,\\
\mbox{ }(3) $\eta \in (3,3)\mbox{-}SAT_{N}$ and $\eta$ is satisfiable,\\
\mbox{ }(4) $\eta \in (3,4)\mbox{-}SAT_{N}$ or \\
\mbox{ }(5) $\eta$ can be reduced to $\tilde \eta \in (3,4)\mbox{-}SAT_{N}$
in polynomial time in the length of $\eta$.\\
Moreover, checking if $\eta \in \cup_{s=1}^3(3,s)\mbox{-}SAT_{N}$ takes 
polynomial time in the length of $\eta$.

\noindent $\it Proof$.
By the definition, $\cup_{s=1}^{\infty}(3,s)\mbox{-}SAT \subset 3SAT$. Since
$(3,4)\mbox{-}SAT$ is $\mathcal{NP}$-complete \cite{Tovey1}, for $s > 4$, any
$\eta \in (3,s)\mbox{-}SAT$ can be reduced to $\tilde \eta \in (3,4)\mbox{-}SAT$
in polynomial time.  We show that this reduction transforms 
$\eta \in (3,s)\mbox{-}SAT_{N}$ to $\tilde\eta \in (3,4)\mbox{-}SAT_{N}$
in polynomial time in the length of $\eta$. Thus, for every instance $\eta$ 
of $3SAT_{N}$, $\eta \in \cup_{s=1}^4(3,s)\mbox{-}SAT_{N}$
or (5) is true.  The second part of (3) is the special case of Theorem 2.4 
\cite{Tovey1}. We just need to prove the second parts of (1) and (2). By
the definition of $(r,s)\mbox{-}SAT_{N}$, the second part of (1) is trivial.
For the second part of (2), from the assumption, instance $\eta$ has at least
three variables, and at least one of them with occurrence number 2, say $x_1,
x_2$ or $x_3$. Let $\tilde \eta = \eta \wedge (x_1 \vee x_2 \vee x_3)$,
then $\tilde \eta \in (3,3)\mbox{-}SAT_{N}$ and $\tilde \eta$ is satisfiable.
Thus, $\eta$ is satisfiable as well.

For the last statement of Theorem 2, suppose that $\eta$ is generated
by variables and their negations 
$\{x_{i_1}, \cdots, x_{i_n}, \neg x_{i_1}, \cdots, \neg x_{i_n}\}$
where $i_1 < \cdots < i_n$. First inspect a variable and its negation
$\{x_{i_1}, \neg x_{i_1}\}$, let $c$ be the number of their occurrences.
If $c > 3$, $\eta \not\in \cup_{s=1}^3(3,s)\mbox{-}SAT_{N}$,
this process completes. Otherwise, inspect $\{x_{i_2}, \neg x_{i_2}\}$,
if the number of their occurrences is greater than $c$, let $c$ be the
number of their occurrences. If $c > 3$, 
$\eta \not\in \cup_{s=1}^3(3,s)\mbox{-}SAT_{N}$, 
this process completes. Otherwise, repeat this
process for $\{x_{i_3}, \neg x_{i_3}\}, \cdots, \{x_{i_n}, \neg x_{i_n}\}$.
If the result $c > 3$, $\eta \not\in \cup_{s=1}^3(3,s)\mbox{-}SAT_{N}$;
if $c \le 3$, $\eta \in \cup_{s=1}^3(3,s)\mbox{-}SAT_{N}$.
Clearly, this process takes polynomial time in the length of $\eta$.

We would like to remark that in the statement (5) above, if $\eta$ has
$m$ clauses, then $\tilde \eta$ has at most $43m$ clauses as shown in 
\cite{Tovey1}. This upper bound can be reduced from $43m$ to $31m$ by
modifying Tovey's procedure as follows:

Step 1. Check if there is any variable with more than 4 occurrences in
the expression. If there is no such variable, the process completes.
Otherwise, go to Step 2.

Step 2. For convenience, we may assume that variable $x$ appears in $k$
clauses with $k > 4$. Other cases can be handled in a similar way.
Create $k$ new variables $x_1, \cdots, x_k$ and replace the $i$th occurrence
of $x$ with $x_i$, $i = 1, \cdots, k$. Create clauses $(x_i \vee \neg x_{i+1})$
for $i = 1, \cdots, k-1$ and clause $(x_k \vee \neg x_1)$. 
The clause $(x_i \vee \neg x_{i+1})$ implies that if $x_i$ is false, $x_{i+1}$
must be false as well. The cyclic structure of the clauses therefore forces the
$x_i$ to be either all true or all false.
For each clause
$(x_i \vee \neg x_{i+1})$, for $i = 1, \cdots, k-1$, and the clause 
$(x_k \vee \neg x_1)$, introduce new variable $y^{(x)}_i$, so that the
clause becomes $(x_i \vee \neg x_{i+1} \vee \neg y^{(x)}_i)$ or
$(x_k \vee \neg x_1 \vee \neg y^{(x)}_k)$. Now note that we can force each
$y^{(x)}_i$ to be true by means of the clauses below in which $y^{(x)}_i$
appears only three times and other variables appear four times:
\begin{eqnarray}
&&(y^{(x)}_i \vee a^{(x)}_{i,j} \vee b^{(x)}_{i,j}), \mbox{ $j = 1, 2,$} \nonumber \\
&&(d^{(x)}_{i,j} \vee \neg a^{(x)}_{i,j} \vee \neg b^{(x)}_{i,j}),
(d^{(x)}_{i,j} \vee \neg a^{(x)}_{i,j} \vee b^{(x)}_{i,j}), 
(d^{(x)}_{i,j} \vee a^{(x)}_{i,j} \vee \neg b^{(x)}_{i,j}), \mbox{ $j = 1, 2,$} \nonumber \\
&&(\neg d^{(x)}_{i,1} \vee \neg d^{(x)}_{i,2} \vee y^{(x)}_i). \nonumber 
\end{eqnarray}
In order to make the above nine clauses satisfiable, $y^{(x)}_i$ must be true.
The proof is the same as the proof in the first case of subroutine (3) of
Theorem 1.
Append the clause $(x_i \vee \neg x_{i+1} \vee \neg y^{(x)}_i)$ for 
$i = 1, \cdots, k-1$, the clause $(x_k \vee \neg x_1 \vee \neg y^{(x)}_k)$
and their associate nine clauses above to the modified expression. Note
that the expression leaving this step is satisfiable if and only if the 
expression entered to this step is satisfiable. Go to Step 1. 

Since Step 2 reduces one variable with occurrences greater than 4 each
time and does not create any more such variable, and the original
expression $\eta$ has finite number of such variables, the process
terminates. Since $\eta$ is normal and the modified expression in
each Step 2 keeps normal, the result expression is normal as well.
Clearly, if $\eta$ has $m$ clauses, the final expression has at most
$m + 3m + 27m = 31m$ clauses, and the procedure transforming any
$\eta \in (3, s)\mbox{-}SAT_{N}$ with $s > 4$ to 
$\tilde \eta \in (3,4)\mbox{-}SAT_{N}$ takes polynomial time in
the length of $\eta$. Now the proof of the Classification Theorem 
completes.

\noindent $\bf Corollary\mbox{ }1$.
$\cup^4_{s=1}(3,s)\mbox{-}SAT_{N}$ is $\mathcal {NP}$-complete.

\noindent $\it Proof$. 
First, we show that $\cup^4_{s=1}(3,s)\mbox{-}SAT_{N}$ is
in $\mathcal{NP}$. A nondeterministic polynomial time Turing
machine can guess a truth assignment to a given expression 
$\eta \in \cup^4_{s=1}(3,s)\mbox{-}SAT_{N}$ and accept if
the assignment satisfies $\eta$. Now Corollary 1 follows
directly from Theorem 1 and Theorem 2.

As in the proof of Theorem 1, we can show that 
$\cup^{\infty}_{n=3}3SAT_{N}(n)$ is in $\mathcal{NP}$.
From Section 2, for any instance $\eta \in 3SAT_{N}$, there 
exists the polynomial time map $\phi_{map}$, such that
$\phi_{map}(\eta) = \tilde\eta \in 3SAT_{N}(n)$
for some integer $n$ which depends on $\eta$, and 
$\tilde\eta$ is satisfiable if and only if $\eta$ is satisfiable. 
We can view $\phi_{map}$ as a polynomial time reduction 
from $3SAT_{N}$ to $\cup^{\infty}_{n=3}3SAT_{N}(n)$.
Moreover, $\phi_{map}$ keeps the normalness from the
definition. Thus, we can view $\phi_{map}$ as the polynomial
time reduction from 
$3SAT_{N}$ to $\cup^{\infty}_{n=3}3SAT_{N}(n)$
as well. Now we have the following:

\noindent $\bf Corollary\mbox{ }2$.
$\cup^{\infty}_{n=3}3SAT_{N}(n)$ is $\mathcal {NP}$-complete.

We would like to give more detail to the Step 1 in the proof 
of Theorem 2, and rewrite it as an algorithm. That is, for any 
\begin{eqnarray}
\eta &=&(y_{k_1} \vee y_{k_2}  \vee y_{k_3}) \wedge 
(y_{k_4} \vee y_{k_5} \vee y_{k_6}) \wedge \cdot \cdot \cdot \wedge \nonumber \\
&&(y_{k_{3m+1}} \vee y_{k_{3m+2}}\vee y_{k_{3m+3}}) \in SAT_{N}(n), \nonumber
\end{eqnarray} 
the algorithm checks if each variable has less than 
4 occurrences in $\eta$.  If so, it returns $true$; otherwise, 
it returns $false$.

\noindent $\bf Algorithm\mbox{ }3$.\\
\indent for $i = 1, 2, \cdots, n$ \\
\indent\indent set $c = 0$;\\
\indent\indent for $j = 1, 2, \cdots, 3(m+1)$ \\
\indent\indent\indent if $e_i(y_{k_j}) = true \mbox{ or } false$, increase $c$  by $1$;\\
\indent\indent\indent if $c > 3$, return $false$;\\
\indent if $i = n$, return $true$.

\noindent The elementary steps of Algorithm 3 are setting $c = 0$, 
increasing $c$ by 1, checking if $c > 3$ and evaluating $e_i(y_{k_j})$,
for $i = 1, 2, \cdots, n$ and $j = 1, 2, \cdots, 3(m+1)$, and returning
$true$ or $false$.  It is a polynomial time algorithm.

\subsection*{5. Aggressive Truth Assignments}
In this section, we introduce the concept of aggressive 
truth assignment and endow the set of all aggressive truth 
assignments with a metric. We use Theorem 2
to extend the definition of generalized truth assignment as follows: 
for any  $\eta \in 3SAT_{N}(n)$,
(1) it evaluates $\eta$ as a truth assignment using Algorithm 1;
(2) it checks if $\eta \in \cup_{s=1}^3(3,s)\mbox{-}SAT_{N}(n)$.
We call this extended generalized truth assignment an aggressive
truth assignment and, for convenience, we continue to use the 
same notation as the truth assignment. If an instance of 
$3SAT_{N}(n)$ 
is in $\cup_{s=1}^3(3,s)\mbox{-}SAT_{N}(n)$, the aggressive 
truth assignment $e_1e_2 \cdots e_m$ just returns a $true$ value. 
For any $\eta \in 3SAT_{N}(n)$, 
the aggressive truth assignment $e_1e_2 \cdots e_m$ works 
in this way:

\noindent (1) It evaluates $\eta$ as a generalized truth assignment
as shown in Algorithm 1. If $e_1e_2 \cdots e_m(\eta)$ = $true$, 
it returns a $true$ value and the process completes, 
otherwise it returns a $false$ value and goes to next 
subroutine.

\noindent (2) It checks if $\eta \in \cup_{s=1}^3(3,s)\mbox{-}SAT_{N}(n)$,
using Algorithm 3.  If so, it returns a $true$ value, otherwise it 
returns a $false$ value. 

\noindent So if $e_1e_2 \cdots e_m(\eta) = true$,  then $\eta$ is 
satisfiable.  In other words, if $e_1e_2 \cdots e_m(\eta) = false$, 
$\eta$ must be an instance of the $\mathcal {NP}$-complete problem. 
The aggressive truth assignments catch all easily decidable 
problems under the sense of the Classification Theorem and decide 
each instance in those problems in polynomial time.

In order to prove that an algorithm defined by a sequence of 
polynomial time algorithms in Section 8 is still a polynomial time 
algorithm, we want to extend the definition of aggressive truth 
assignment as follows.  A generalized truth assignment is called less efficient 
aggressive if for any instance $\eta \in 3SAT_{N}(n)$,
(1) it evaluates $\eta$ as a truth assignment using Algorithm 2;
(2) it checks if $\eta \in \cup_{s=1}^3(3,s)\mbox{-}SAT_{N}(n)$.
For any $\eta \in 3SAT_{N}(n)$, the less efficient aggressive truth 
assignment $e_1e_2 \cdots e_m$ works in this way:

\noindent (1) It evaluates $\eta$ as a generalized truth assignment 
as shown in Algorithm 2. If $e_1e_2 \cdots e_m(\eta)$ = $0$, 
it returns a $true$ value and the process completes, otherwise it 
returns a $false$ value and goes to next subroutine.

\noindent (2) It checks if $\eta \in \cup_{s=1}^3(3,s)\mbox{-}SAT_{N}(n)$,
using Algorithm 3.  If so, it returns a $true$ value, otherwise it 
returns a $false$ value. 

For any two aggressive truth assignments or less efficient 
aggressive truth assignments 
$e_1^{1}e_2^{1} \cdots e_m^{1}$
and
$e_1^{2}e_2^{2} \cdots e_m^{2}$, we define the composition
$\varphi = e_1^{1}e_2^{1} \cdots e_m^{1}\cdot
e_1^{2}e_2^{2} \cdots e_m^{2}$ as follows:
for any $\eta \in 3SAT_{N}(n)$,

\noindent (a) if $e_1^{2}e_2^{2} \cdots e_m^{2}(\eta) = true$,
$\varphi(\eta) = true$ and skips the evaluation 
$e_1^{1}e_2^{1} \cdots e_m^{1}(\eta)$;

\noindent (b) if $e_1^{2}e_2^{2} \cdots e_m^{2}(\eta) = false$ and
$e_1^{1}e_2^{1} \cdots e_m^{1}(\eta) = true$, $\varphi(\eta) = true$;

\noindent (c) if $e_1^{2}e_2^{2} \cdots e_m^{2}(\eta) = false$ and
$e_1^{1}e_2^{1} \cdots e_m^{1}(\eta) = false$,
$\varphi(\eta) = false$.

\noindent Thus, from the definition, if $\varphi(\eta) = true$, $\eta$ is
satisfiable.  We can extend this definition to the composition of k 
aggressive truth assignments or less efficient aggressive truth 
assignments for $k \geq 2$. 

Let
\begin{eqnarray}
{\mathcal{TA}}^1 &=& \{ a \mid a \mbox{ is an aggressive truth assignment} \} \nonumber \\
{\mathcal{TA}}^k &=& \{ (a_1)( a_2) ... (a_k) \mid a_1,  a_2,  ...,  a_k \in {\mathcal{TA}}^1\},  \mbox{ where $k \ge 2$,  and }
 \label{eqn5.3} \\
{\mathcal{TA}}^{\infty} &=& \cup_{k=1}^{\infty}{\mathcal{TA}}^k.\nonumber
\end{eqnarray}
We introduce a metric in ${\mathcal{TA}}^{\infty} \times {\mathcal{TA}}^{\infty}$.
The distance between two atomic truth assignments is defined as: 
$d(x^{\ast}_i, \neg x^{\ast}_j) =$ $d(\neg x^{\ast}_j, x^{\ast}_i)$ $= \frac{i+j}{2^{i+j+2}}$,
and 
$d(x^{\ast}_i, x^{\ast}_j) = d(x^{\ast}_j, x^{\ast}_i) = 
d(\neg x^{\ast}_i, \neg x^{\ast}_j) = d(\neg x^{\ast}_j, \neg x^{\ast}_i) =
\frac{|i-j|}{2^{i+j+2}}$ for all integers $i,j \ge 1$. 
Thus, $d(x^{\ast}_i, \neg x^{\ast}_i) =$ $d(\neg x^{\ast}_i, x^{\ast}_i)$ $= \frac{i}{2^{2i+1}}$,
and $d(e_i, e_i) = 0$, for all integers $i \ge 1$.
For any $e_1 e_2 \cdots e_m, e_1^{\prime} e_2^{\prime} \cdots e_m^{\prime}  \in {\mathcal{TA}}^1$,
it is defined as
\begin{eqnarray}
d(e_1 e_2 \cdots e_m, e_1^{\prime}e_2^{\prime} \cdots e_m^{\prime}) = 
\sum_{k=1}^\infty d(e_k,e_k^{\prime}).
\label{eqn5.4}
\end{eqnarray}
From the definition, for any $a_1, a_2, a_3  \in {\mathcal{TA}}^1$, we have
\begin{eqnarray}
&&d(a_1, a_2) \ge 0, \nonumber \\
&&d(a_1, a_2) = 0 \mbox{ if and only if } a_1 = a_2, \nonumber \\
&&d(a_1, a_2) = d(a_2, a_1), \mbox{ and} \nonumber \\
&&d(a_1, a_3) \le d(a_1, a_2) + d(a_2, a_3).\nonumber
\end{eqnarray}
So $d: {\mathcal{TA}}^1 \times {\mathcal{TA}}^1 \rightarrow [0, \infty)$ is
a metric and $({\mathcal{TA}}^1, d)$ is a metric space.
The above definitions can be extended to the following cases:
$d(e_i, \cdot) = d(\cdot, e_i) = \frac{i}{2^{i+2}}$ for all integers $i \ge 1$
where $\cdot$ is the empty parameter.
For any $e_1 e_2 \cdots e_m \in {\mathcal{TA}}^1$, 
$d(e_1 e_2 \cdots e_m, \cdot) = d(\cdot, e_1 e_2 \cdots e_m) = \sum_{k=1}^\infty d(e_k, \cdot)
= \sum_{k=1}^\infty \frac{k}{2^{k+2}}$ where $\cdot$ is the empty parameter.
For any $a_1, \cdots, a_m, b_1, \cdots, b_n \in {\mathcal{TA}}^1$, 
\begin{eqnarray}
d(a_1 \cdots a_m, b_1 \cdots b_n) =
\left\{ \begin{array}{lllll}
\frac{1}{1^2}d(a_1, b_1) + \cdots + \frac{1}{m^2}d(a_m, b_m) + \\
\mbox{   }\frac{1}{(m+1)^2}d(b_{m+1}, \cdot)\mbox{ }+ \cdots + \frac{1}{n^2} d(b_{n}, \cdot) & \mbox{ if $0 < m < n$,} \\
\frac{1}{1^2}d(a_1, b_1) + \cdots + \frac{1}{n^2}d(a_n, b_n) + \\
\mbox{   }\frac{1}{(n+1)^2}d(a_{n+1}, \cdot)\mbox{ }+ \cdots + \frac{1}{m^2} d(a_{m}, \cdot) & \mbox{ if $0 < n < m $, and} \\
\frac{1}{1^2}d(a_1, b_1) + \cdots + \frac{1}{n^2}d(a_n, b_n) & \mbox{ if $0 < m = n$.}
\end{array} \right.  
\label{eqn5.5}
\end{eqnarray}
From the definition, we can verify the following:

\noindent $\bf Lemma\mbox{ }2$.
$d: \mathcal{TA}^{\infty} \times \mathcal{TA}^{\infty} \rightarrow [0, \infty)$ is
a metric and $(\mathcal{TA}^{\infty}, d)$ is a metric space.

\subsection*{6. Pseudo-algorithms}
Suppose that there are some polynomial time algorithms on 
$3SAT_{N}$. They are polynomial time algorithms on 
$\cup_{n=3}^{\infty}3SAT_{N}(n)$ as well. Let
\begin{eqnarray}
&\empty&\mathcal{A} = \{ f_\xi \mid f_\xi
\mbox{ is a polynomial time algorithm on }\cup_{n=3}^{\infty}3SAT_{N}(n)\}. \label{eqn6.1}
\end{eqnarray}
We prove that $\mathcal A$ is empty using the proof by contradiction in 
the following sections. Suppose that $\mathcal A$ is not empty. 

For each aggressive truth assignment $a \in {\mathcal{TA}}^1$, we define
\begin{eqnarray}
\mathcal A a = \{ f_\xi a \mid f_\xi \in \mathcal A \},
\label{eqn6.2}
\end{eqnarray}
and $f_\xi a$ as follows
\begin{eqnarray}
 f_\xi a(\eta) = \left\{ \begin{array}{ll}
 true & \mbox{if $a(\eta) = true$},\\ 
 f_\xi (\eta) & \mbox{if $a(\eta) \not= true$},\end{array} \right.  
\label{eqn6.3}
\end{eqnarray}
for any instance $\eta \in 3SAT_{N}(n)$ and $n \ge 3$.  If $a(\eta)$ $= true$, 
$\eta$ is satisfiable and the algorithm terminates. If $a(\eta) \not= true$, 
the algorithm applies $f_\xi$ to $\eta$.  Note that, for any 
$a,  b \in {\mathcal{TA}}^1$, concerning not only final results but also 
computation processes or steps, we have  $f_\xi a \not= f_\xi a a$,
if $a \not=  b$, then $f_\xi a b \not= f_\xi b a$.

Suppose that $\mathcal{A}a \subset \mathcal{A}$ for any $a \in {\mathcal{TA}}^1$.
Choose an arbitrary $f \in \mathcal{A}$.
Define the following sets:
\begin{eqnarray}
< f >^0 &=& \{ f\} \nonumber \\
< f >^k &=& \{f \alpha \mid \alpha \in \mathcal{TA}^k\} \mbox{ for } k = 1, 2, \cdots \mbox{ and,} \label{eqn6.4} \\
< f > &=& \cup_{k=0}^{\infty}<f>^k.\nonumber
\end{eqnarray}
We have $< f >^k \bigcap < f >^l = \emptyset$ if $k \not= l$
and $< f >^k \not= \emptyset$ by the definition, and
$< f >^k \subset \mathcal{A}$ for all integers $k \ge 0$ by the assumption.

For any $a_1, \cdots, a_m, b_1, \cdots, b_n \in {\mathcal{TA}}^1$, 
and $f a_1 \cdots a_m$, $f b_1 \cdots b_n \in < f >$, define 
\begin{eqnarray}
d(f a_1 \cdots a_m, f b_1 \cdots b_n) =
\left\{ \begin{array}{ll}
0 & \mbox{  if $m = n = 0$}, \\
d(a_1 \cdots a_m, b_1 \cdots b_n) & \mbox{ if $m > 0$ or $n > 0$.}
\end{array} \right.  
\label{eqn6.5}
\end{eqnarray}
From Lemma 2, we can see that $d: < f > \times < f > \mbox{} \rightarrow [0, \infty)$ 
is a metric and $(< f >, d)$ is a metric space. 
For any $f\alpha_1 \in < f >^k$, $f\alpha_2 \in < f >^l$ and $\beta \in \mathcal{TA}^m$
with $m \ge 1$,
if $k \not= l$, $d(f\alpha_1 \beta, f\alpha_2 \beta) < d(f\alpha_1, f\alpha_2)$; if $k = l$, 
$d(f\alpha_1 \beta, f\alpha_2 \beta) = d(f\alpha_1, f\alpha_2)$. Thus, for any 
$\beta \in \mathcal{TA}^m$, the map $\beta: < f >^k \rightarrow < f >^{k+m}$ 
is equidistant. The metric $d$ can be extended from $< f > \times < f >$ to 
$\mathcal{A} \times \mathcal{A}$ as follows:  For any given
$g_1, g_2 \in \mathcal{A}$,  if $g_1 \not\in < f >$ or $g_2 \not\in < f >$   
for any $f \in \mathcal{A}$, define $d(g_1, g_2) = 1$.  For such 
$g_1$ and $g_2$,  for any $\beta \in \mathcal{TA}^m$ with $m \ge 1$, 
$d(g_1\beta, g_2\beta) = d(g_1, g_2) = 1$. 

A sequence $\{f_n\}$ of $< f >$ is called Cauchy if for any real number 
$\epsilon > 0$, there exists an integer $N > 0$, such that for all natural
numbers $m, n > N$, $d(f_m, f_n) < \epsilon$. 

\noindent $\bf Definition\mbox{ }1.$
A set $\mathcal{S}$ is called compatible with the $\mathcal P$ versus $\mathcal {NP}$ problem
on $\cup_{n=3}^{\infty}3SAT_{N}(n)$ if it satisfies:\\
(1) for any $\rho \in \mathcal{S}$ and $\eta \in \cup_{n=3}^{\infty}3SAT_{N}(n)$, 
$\rho$ checks if $\eta \in \cup_{s=1}^3\cup_{n=3}^{\infty}(3,s)\mbox{-}SAT_{N}(n)$;\\
(2)  $\rho$ is not an algorithm on $\cup_{n=3}^{\infty}3SAT_{N}(n)$;\\ 
(3) for any algorithm $f_\xi$ on $\cup_{n=3}^{\infty}3SAT_{N}(n)$,
 $f_\xi\rho$ is an algorithm on the same set;\\
(4) some elements of $\mathcal{S}$ can generate 
an algorithm on $\cup_{n=3}^{\infty}3SAT_{N}(n)$ whose time complexity\\
\indent exceeds polynomial however.

\noindent $\bf Definition\mbox{ }2.$ 
Any element of $\mathcal{S}$ is called a pseudo-algorithm on $\cup_{n=3}^{\infty}3SAT_{N}(n)$.

\noindent $\bf Lemma \mbox{ }3$. 
For any integer $n \ge 3$ and $a \in \mathcal{TA}^{1}$, there exists an 
$\eta \in 3SAT_N(n)$ such that $a(\eta) = true$ and $a^{\prime}(\eta) = false$
for any $a^{\prime} \in \mathcal{TA}^{1}$ which is not identical with $a$ 
in the first $n$ atomic truth assignments.    

\noindent $\it Proof$. Suppose that $a = e_1e_2 \cdots e_n$. 
Note that the following evaluations:
$e_2(x_2) \vee e_3(x_3)$, $e_2(x_2) \vee e_3(\neg x_3)$,
$e_2(\neg x_2) \vee e_3(x_3)$ and $e_2(\neg x_2) \vee e_3(\neg x_3)$
always have one and only one $true$ value which does not rely on $e_2 = x_2^{\ast}$ 
or $e_2 = \neg x_2^{\ast}$ and $e_3 = x_3^{\ast}$ or $e_3 = \neg x_3^{\ast}$.
If $e_1 = x_1^{\ast}$, let 
\begin{eqnarray}
\eta_1 &=& (x_1 \vee x_2 \vee x_3) \wedge (x_1 \vee x_2 \vee \neg x_3) \wedge \nonumber \\ 
&&(x_1 \vee \neg x_2 \vee x_3)  \wedge (x_1 \vee \neg x_2 \vee \neg x_3); 
\label{eqn6.6}
\end{eqnarray}
if $e_1 = \neg x_1^{\ast}$, let
\begin{eqnarray} 
\eta_1 &=& (\neg x_1 \vee x_2 \vee x_3) \wedge (\neg x_1 \vee x_2 \vee \neg x_3) \wedge \nonumber \\ 
&&(\neg x_1 \vee \neg x_2 \vee x_3) \wedge (\neg x_1 \vee \neg x_2 \vee \neg x_3).
\label{eqn6.7}
\end{eqnarray}
The result of evaluation of $\eta_1$ by any truth assignment is determined by 
$e_1 = x_1^{\ast}$ or $e_1 = \neg x_1^{\ast}$ only. From (\ref{eqn6.6}) 
and (\ref{eqn6.7}), we can see that $e_1e_2e_3(\eta_1) = true$ and 
$\neg e_1 e_2e_3(\eta_1) = false$ which does not rely on 
$e_2 = x_2^{\ast}$ or $e_2 = \neg x_2^{\ast}$ and 
$e_3 = x_3^{\ast}$ or $e_3 = \neg x_3^{\ast}$. Let $\eta = \eta_1$.
If $n > 3$, we can extend this construction to the following groups of 
variables: $(x_2, x_3, x_4)$, $\cdots, (x_{n-2}, x_{n-1}, x_n)$, 
$(x_{n-1}, x_n, x_1)$ and $(x_n, x_1, x_2)$;  if $n = 3$, we just
put two more groups of variables $(x_2, x_3, x_1)$ and $(x_3, x_1, x_2)$ 
to the construction list. We then wedge the associated wedge of 
clauses $\eta_2, \cdots, \eta_n$ to $\eta$ in both cases.  For each 
group of these variables, for example $(x_k, x_{k+1}, x_{k+2})$, 
if $e_k = x_k^{\ast}$, let 
\begin{eqnarray}
\eta_k &= &(x_k \vee x_{k+1} \vee x_{k+2}) \wedge (x_k \vee x_{k+1} \vee \neg x_{k+2}) \wedge \nonumber \\  
&&(x_k \vee \neg x_{k+1} \vee x_{k+2}) \wedge (x_k \vee \neg x_{k+1} \vee \neg x_{k+2});
\label{eqn6.8}
\end{eqnarray}
if $e_k = \neg x_k^{\ast}$, let
\begin{eqnarray}
\eta_k &=&(\neg x_k \vee x_{k+1} \vee x_{k+2}) \wedge (\neg x_k \vee x_{k+1} \vee \neg x_{k+2}) 
\wedge \nonumber \\  
&&(\neg x_k \vee \neg x_{k+1} \vee x_{k+2}) \wedge (\neg x_k \vee \neg x_{k+1} \vee \neg x_{k+2}).
\label{eqn6.9}
\end{eqnarray}
The result of evaluation of $\eta_k$ by any truth assignment is determined 
by $e_k = x_k^{\ast}$ or $e_k = \neg x_k^{\ast}$ only.  We can verify from 
(\ref{eqn6.8}) and (\ref{eqn6.9}) that $\eta_k$ has the following property: 
$e_k e_{k+1} e_{k+2}(\eta_k) = true$ and $\neg e_k e_{k+1} e_{k+2}(\eta_k) = false$
which does not rely on $e_{k+1} = x_{k+1}^{\ast}$ or $e_{k+1} = \neg x_{k+1}^{\ast}$ and 
$e_{k+2} = x_{k+2}^{\ast}$ or $e_{k+2} = \neg x_{k+2}^{\ast}$.
From the above construction,  $\eta = \eta_1 \wedge \eta_2 \wedge \cdots \wedge \eta_n$.
Putting all properties of $\eta_1, \eta_2, \cdots, \eta_n$ together, we can see that 
$a(\eta) = true$ and $a^{\prime}(\eta) = false$ for any $a^{\prime}$ 
that is not identical with $a$ in the first $n$ atomic truth assignments.
The proof of Lemma 3 completes.

\noindent $\bf Example\mbox{ }2$. 
Let $a^{\ast}$ be the negative aggressive truth assignment, i.e. 
$a^{\ast} = \neg x_1^{\ast} \neg x_2^{\ast} \cdots \neg x_n^{\ast} \cdots$.
For each $n \ge 3$, the expression $\eta \in 3SAT_N(n)$ defined as 
\begin{eqnarray}
\eta &=& (\neg x_1 \vee x_2 \vee x_3) \wedge (\neg x_1 \vee x_2 \vee \neg x_3) \wedge \nonumber \\ 
&&(\neg x_1 \vee \neg x_2 \vee x_3)  \wedge (\neg x_1 \vee \neg x_2 \vee \neg x_3) \wedge \nonumber \\
&&\cdots \cdots \nonumber \\
&& (\neg x_{n-1} \vee x_n \vee x_1) \wedge (\neg x_{n-1} \vee x_n \vee \neg x_1) \wedge \nonumber \\ 
&&(\neg x_{n-1} \vee \neg x_n \vee x_1)  \wedge (\neg x_{n-1} \vee \neg x_n \vee \neg x_1) \wedge \nonumber \\
&& (\neg x_n \vee x_1 \vee x_2) \wedge (\neg x_n \vee x_1 \vee \neg x_2) \wedge \nonumber \\ 
&&(\neg x_n \vee \neg x_1 \vee x_2)  \wedge (\neg x_n \vee \neg x_1 \vee \neg x_2)  
\label{eqn6.10}
\end{eqnarray}
has the property of Lemma 3.  That is $a^{\ast}(\eta) = true$ and  
$a^{\prime}(\eta) = false$ for any $a^{\prime} \in \mathcal{TA}^{1}$ 
which is not identical with $a$ in the first $n$ atomic truth assignments.

\noindent $\bf Proposition\mbox{ }1.$  $\mathcal{AT}^{1}$ is compatible 
with the $\mathcal P$ versus $\mathcal {NP}$ problem on 
$\cup_{n=3}^{\infty}3SAT_{N}(n)$.

We postpone the proof of Proposition 1 in Section 9. One corollary of 
Proposition 1 is that any aggressive truth assignment is a pseudo-algorithm 
on $\cup_{n=3}^{\infty}3SAT_{N}(n)$.

For any algorithm
$\varphi$ on $\cup_{n=3}^{\infty}3SAT_{N}(n)$,
we define
\begin{eqnarray}
\mathcal A \varphi = \{ f_\xi \varphi \mid f_\xi \in \mathcal A \},
\label{eqn6.11}
\end{eqnarray}
and $f_\xi \varphi$ as follows
\begin{eqnarray}
 f_\xi \varphi(\eta) = \left\{ \begin{array}{ll}
 true & \mbox{if $\varphi(\eta) = true$},\\
 false & \mbox{if $\varphi(\eta) \not= true$}, \end{array} \right.  
\label{eqn6.12}
\end{eqnarray}
for any $\eta \in \cup_{n=3}^{\infty}3SAT_{N}(n)$, i.e., 
$f_\xi \varphi \equiv \varphi$ on $\cup_{n=3}^{\infty}3SAT_{N}(n)$.

\subsection*{7. Equivalence Classes}
In practice, when  implementing an algorithm, we usually break 
it to some processes.  Each process has its subroutines, and each 
subroutine has its steps. Depending on the complexity of the 
algorithm, we may break it to more or less levels. For a given 
algorithm, if the inputs are the same, the outputs or the results are 
the same, and the implementations have the same sequence of 
steps. This sequence of steps is called an implementation sequence. 
We handle any pseudo-algorithm in a similar way.

\noindent $\bf Definition\mbox{ }3.$ A bijective map 
$\pi$ from $\cup_{n=3}^{\infty}3SAT_{N}(n)$ to itself
is ordered if $\pi(3SAT_{N}(n)) = 3SAT_{N}(n)$ for all 
integers $n \ge 3$.

\noindent $\bf Definition\mbox{ }4.$  Algorithms $\varphi_1$ and 
$\varphi_2$ on $\cup_{n=3}^{\infty}3SAT_{N}(n)$
are equivalent if there exists a bijective and ordered map $\pi$ from 
$\cup_{n=3}^{\infty}3SAT_{N}(n)$
to itself, such that for any expression
$\eta \in \cup_{n=3}^{\infty}3SAT_{N}(n)$,
$\varphi_1(\eta)$ and $\varphi_2(\pi(\eta))$ have the same implementation 
sequences.

\noindent $\bf Definition\mbox{ }5.$ Pseudo-algorithms $\rho_1$ and 
$\rho_2$ on $\cup_{n=3}^{\infty}3SAT_{N}(n)$
are equivalent if there exists a bijective and ordered map 
$\pi$ from $\cup_{n=3}^{\infty}3SAT_{N}(n)$
to itself, such that for any expression  
$\eta \in \cup_{n=3}^{\infty}3SAT_{N}(n)$,
$\rho_1(\eta)$ and $\rho_2(\pi(\eta))$ have the same implementation 
sequences.

If steps $s_1$ and $s_2$ are the same, we write $s_1 = s_2$. Suppose 
that $\sigma_1$ and $\sigma_2$ are algorithms or pseudo-algorithms 
on $\cup_{n=3}^{\infty}3SAT_{N}(n)$.
If $\sigma_1$ and $\sigma_2$ are equivalent, we write 
$\sigma_1 \equiv \sigma_2$. The following remarks set up the rule to 
compare two implementation sequences of some pseudo-algorithms.

\noindent $\bf Remark\mbox{ }2.$  In Algorithm 1, the following equations
give all the same steps: \\
\noindent (1) $e_i(y_i) = e_i(y_i)$ and $\neg e_i(y_i) = \neg e_i(y_i)$ where 
$y_i = x_i \mbox{ or } \neg x_i$, for $i = 1, 2, \cdots, n$;\\
\noindent (2) $e_i(y_i) = \neg e_i(\neg y_i)$
where $y_i = x_i \mbox{ or } \neg x_i$, for $i = 1, 2, \cdots, n$;\\
\noindent (3) $e_i(x_j) = e_i(\neg x_j) = \neg e_i(x_j) = \neg e_i(\neg x_j)$ if $i \not= j$;\\
\noindent (4) $\{\mbox{returning } true\} = \{\mbox{returning } true\}$ and
$\{\mbox{returning } false\} = \{\mbox{returning } false\}$.  

\noindent $\bf Remark\mbox{ }3.$  In Algorithm 3, the following equations
give all the same steps: \\
\noindent (1) $\{\mbox{setting } c = 0 \} = \{\mbox{setting } c = 0 \}$ and
$\{\mbox{increasing } c  \mbox{ by } 1 \} = \{\mbox{increasing } c  \mbox{ by } 1 \}$;\\
\noindent (2) $\{\mbox{checking if } c > 3 \} = \{\mbox{checking if } c > 3 \}$;\\
\noindent (3) $e_i(y_i) = e_i(y_i)$ and $\neg e_i(y_i) = \neg e_i(y_i)$ 
where $y_i = x_i \mbox{ or } \neg x_i$, for $i = 1, 2, \cdots, n$;\\
\noindent (4) $e_i(y_i) = \neg e_i(\neg y_i)$ 
where $y_i = x_i \mbox{ or } \neg x_i$, for $i = 1, 2, \cdots, n$;\\
\noindent (5) $e_i(x_j) = e_i(\neg x_j) = \neg e_i(x_j) = \neg e_i(\neg x_j)$ if $i \not= j$;\\
\noindent (6) $\{\mbox{returning } true\} = \{\mbox{returning } true\}$ and
$\{\mbox{returning } false\} = \{\mbox{returning } false\}$.

\noindent $\bf Example\mbox{ }3.$ (1) The implementation sequence for
$e_1 e_2 e_3e_4(\eta_1)$ in Example 1 is the following: $e_1(\neg  x_1)$, 
$e_1(\neg x_2)$, $e_2(\neg x_2)$, $e_1(\neg x_1)$, $e_1(x_2)$, $e_2(x_2)$,
$e_2(x_3)$, $e_3(x_3)$ and returning $false$.

\noindent (2) The implementation sequence for checking if 
$\eta_1 \in \cup_{s=1}^3\cup_{n=3}^{\infty}(3,s)\mbox{-}SAT_{N}(n)$        
in Example 1 is the following: $c = 0$,
$e_1(\neg x_1)$, increasing $c$ by 1, checking if $c > 3$, $e_1(\neg x_2)$, 
$e_1(\neg x_3)$, $e_1(\neg x_1)$, increasing $c$ by 1, checking if $c > 3$,
$e_1(x_2)$, $e_1(x_3)$, $c = 0$, $e_2(\neg x_1)$, $e_2(\neg x_2)$,
increasing $c$  by 1, checking if $c > 3$, $e_2(\neg x_3)$, $e_2(\neg x_1)$, 
$e_2(x_2)$, increasing $c$  by 1, checking if $c > 3$, $e_2(x_3)$, $c = 0$, 
$e_3(\neg x_1)$, $e_3(\neg x_2)$, $e_3(\neg x_3)$, increasing $c$  by 1, 
checking if $c > 3$, $e_3(\neg x_1)$, $e_3(x_2)$, $e_3(x_3)$, increasing
$c$  by 1, checking if $c > 3$, $c = 0$, $e_4(\neg x_1)$, $e_4(\neg x_2)$,
$e_4(\neg x_3)$, $e_4(\neg x_1)$, $e_4(x_2)$, $e_4(x_3)$ and 
returning $true$.

The equivalence of algorithms or pseudo-algorithms defined above is 
essentially a special case of Definition 3.2 in \cite{Gurevich1}. 
However, for the different purposes, the equivalence in \cite{Gurevich1}  
is much finer than ones in this paper.

\noindent $\bf Proposition\mbox{ }2.$  Any $a_1, a_2  \in \mathcal{TA}^1$ are
equivalent.

\noindent $\it Proof$.  Let $a_1 = e_1^1 e_2^1 \cdots e_n^1 \cdots$ and 
$a_2 = e_1^2 e_2^2 \cdots e_n^2 \cdots$. Define a map $\pi(a_1) = a_2$
as follows: $\pi(e_i^1) = e_i^2$ for all integers $i \ge 1$. Extend
$\pi$ to from $\cup_{n=3}^{\infty}3SAT_{N}(n)$
to itself as follows: 
\begin{eqnarray}
\pi(y_i) = \left\{ \begin{array}{ll}
 y_i & \mbox{if $\pi(e_i^1) = e_i^1$},\\
 \neg y_i & \mbox{if $\pi(e_i^1) = \neg e_i^1$}, \end{array} \right.  
\label{eqn7.1}
\end{eqnarray}
where $y_i = x_i \mbox{ or } \neg x_i$, and
\begin{eqnarray}
\pi(*) = \left\{ \begin{array}{llll}
 ( & \mbox{if $*$ = (},\\
 ) & \mbox{if $*$ = )},\\
\vee & \mbox{if $* = \vee$},\\
\wedge & \mbox{if $* = \wedge$}.
\end{array} \right.  
\label{eqn7.2}
\end{eqnarray}
It is easy to verify that $\pi$ is a bijective and 
ordered map from 
$\cup_{n=3}^{\infty}3SAT_{N}(n)$
to itself and $\pi^2$ is the identical map.

From the construction of map $\pi$, Remark 2 and Remark 3, for 
$\eta \in 3SAT_{N}(n)$, evaluating $\eta$ 
and $\pi(\eta)$ respectively, $a_1$ and $a_2$ have the same 
implementation sequence, and checking if 
$\eta$ and $\pi(\eta) \in\cup_{s=1}^3(3,s)\mbox{-}SAT_{N}(n)$
respectively, $a_1$ and $a_2$ also have the same implementation 
sequence, i.e., $a_1(\eta)$ and $a_2(\pi(\eta))$ have the same 
implementation sequence. Thus, $a_1 \equiv a_2$. 

The map $\pi$ in Proposition 2 is uniquely determined by $a_1$ and 
$a_2$. On the other hand, from Definition 4, Remark 2 and 
Remark 3, any bijective and ordered map $\pi^{\prime}$ which 
makes $a_1$ and $a_2$ equivalent is identical to $\pi$, i.e., 
we have the following:

\noindent $\bf Lemma\mbox{ }4$.  For any $a_1, a_2 \in \mathcal{TA}^1$,
if the bijective and ordered map $\pi^{\prime}$ makes $a_1$ and 
$a_2$ equivalent, then $\pi^{\prime}$ is identical to $\pi$ which is 
defined in the proof of Proposition 2.

\noindent $\it Proof$.  We use proof by contradiction here.  
Let $a_1 = e_1^1 e_2^1 \cdots e_n^1 \cdots$ and 
$a_2 = e_1^2 e_2^2 \cdots e_n^2 \cdots$ be equivalent 
under $\pi$ and $\pi^{\prime}$.  We may assume that 
$\pi^{\prime}(y_i) = y_i$  or $\pi^{\prime}(y_i) = \neg y_i$,
where $y_i = x_i \mbox{ or } \neg x_i$, for $i = 1, 2, \cdots$. Suppose
that $\pi \not= \pi^{\prime}$. There exists the minimum integer $i_0$ such
that $\pi(x_{i_0}) \not= \pi^{\prime}(x_{i_0})$. We can choose 
$\eta_0 \in 3SAT_{N}(n)$ for some integer $n$,
such that the first clause of $\eta_0$ is $(x_{i_0} \vee y_l \vee y_k)$ where
$y_l = x_l \mbox{ or } \neg x_l$ and $y_k = x_k \mbox{ or } \neg x_k$.
Since $a_1$ and $a_2$ are equivalent under $\pi$ and $\pi^{\prime}$,
$e^1_{i_0}(x_{i_0}) = e^2_{i_0}(\pi(x_{i_0}))$ and 
$e^1_{i_0}(x_{i_0}) = e^2_{i_0}(\pi^{\prime}(x_{i_0}))$.
From Remark 2, $\pi(x_{i_0}) =  \pi^{\prime}(x_{i_0})$,
this contradicts the assumption $\pi(x_{i_0}) \not= \pi^{\prime}(x_{i_0})$.
Thus, we have $\pi^{\prime} = \pi$.

The map $\pi$ making $a_1$ and $b_1$ equivalent is uniquely determined
by $a_1$ and $b_1$. 
Let $a_1 = e_1 \cdots e_{k-1} e_k e_{k+1} \cdots$ and 
$a_1^{\prime} = e_1 \cdots e_{k-1} \neg e_k e_{k+1} \cdots$,
then $a_1^{\prime}$
and $b_1$ are equivalent under different map $\pi^{\prime}$.  
It is interesting to know if  the assumption 
$a_1 \equiv b_1$  and $a_2 \equiv b_2$ under the same map $\pi$ can 
imply that $a_1 = a_2$ and $b_1 = b_2$.  In general, it cannot imply
that. For example, let
$a_1 = x^{\ast}_1 \neg x^{\ast}_2 e_3 \cdots$,
$a_2 = \neg x^{\ast}_1 x^{\ast}_2 e_3 \cdots$,
$b_1 = \neg x^{\ast}_1 x^{\ast}_2 e_3 \cdots$ and
$b_2 = x^{\ast}_1 \neg x^{\ast}_2 e_3 \cdots$.
From the proof of Proposition 2, $a_1 \equiv b_1$ under map $\pi$, where
$\pi(x_1) = \neg x_1$, $\pi(x_2) = \neg x_2$ and $\pi(x_i) = x_i$,
for $i = 3, 4, \cdots$. However, $a_2 \equiv b_2$ under the same map 
$\pi$, $a_1 \not= a_2$ and $b_1 \not= b_2$.  

\noindent $\bf Lemma\mbox{ }5.$  
For any $a_1, a_2, \cdots, a_m, b_1, b_2, \cdots, b_m \in \mathcal{TA}^1$
with $m \ge 2$, $a_1 a_2 \cdots a_m \equiv b_1 b_2 \cdots b_m$ 
 if and only if $a_i \equiv b_i$ under the same map $\pi$, for 
$i = 1, 2, \cdots, m$.

\noindent $\it Proof$.  Suppose that $a_i \equiv b_i$ under the same map 
$\pi$. From the definition of composition of aggressive truth assignments 
and Definition 5, for any  $\eta \in 3SAT_{N}(n)$,
$a_1 a_2 \cdots a_m(\eta)$ and $b_1 b_2 \cdots b_m(\pi(\eta))$
have the same implementation sequences.  Thus,
$a_1 a_2 \cdots a_m \equiv b_1 b_2 \cdots b_m$
under map $\pi$.

Suppose that $a_1 a_2 \cdots a_m \equiv b_1 b_2 \cdots b_m$ under 
map $\pi$. We want to prove that $a_i \equiv b_i$ under the same map 
$\pi$, for $i = 1, 2, \cdots, m$. By Definition 5 and the assumption,
$a_m \equiv b_m$ under the same map $\pi$.  For any $\eta \in 3SAT_{N}(n)$
with $a_m(\eta) = b_m(\pi(\eta)) = true$,  since $a_m$ and $b_m$ are generalized
truth assignments, we can choose some clauses $\theta_1, \cdots, \theta_i$
such that 
$\eta^{\prime} =  \eta \wedge \theta_1 \wedge \cdots \wedge \theta_i\in 3SAT_{N}(n^{\prime})$ 
with $n^{\prime} \ge n$ and $a_m(\eta^{\prime}) = b_m(\pi(\eta^{\prime})) = false$.
Without loss of generality, we may assume that $a_m(\eta) = b_m(\pi(\eta)) = false$.
By Definition 5 and the assumption, $a_{m-1}(\eta)$ and $b_{m-1}(\pi(\eta))$
have the same implementation sequences, so $a_{m-1} \equiv b_{m-1}$ under 
map $\pi$.  We can apply this argument to $k$ and get $a_k \equiv b_k$ 
under the same map $\pi$ for $k = m-2, m - 3, \cdots, 1$. The proof is complete. 

\noindent $\bf Corollary\mbox{ }3.$ 
 For any $a_1, a_2, \cdots, a_m \in \mathcal{TA}^1$ with $m \ge 1$,
$a_1 a_2 \cdots a_m \equiv a_1 a_2 \cdots a_m$ under the identical map. 

\noindent $\it Proof$. It follows directly from Lemma 5.

\noindent $\bf Corollary\mbox{ }4.$ 
 For any $a_1, a_2, \cdots, a_m, b_1, b_2, \cdots, b_m \in \mathcal{TA}^1$ 
with $m \ge 1$,  if $a_1 a_2 \cdots a_m \equiv b_1 b_2 \cdots b_m$ under 
map $\pi$, then $b_1 b_2 \cdots b_m \equiv a_1 a_2 \cdots a_m$ under 
map $\pi^{-1}$.

\noindent $\it Proof$. Since $\pi^2$ is the identical map, $\pi^{-1} = \pi$.
 It follows from Lemma 5.

\noindent $\bf Corollary\mbox{ }5.$   For any 
$a_1, a_2, \cdots, a_m, b_1, b_2, \cdots, b_m, c_1, c_2, \cdots, c_m \in \mathcal{TA}^1$ 
with $m \ge 1$,  if $a_1 a_2 \cdots a_m \equiv b_1 b_2 \cdots b_m$ under 
map $\pi_1$, and $b_1 b_2 \cdots b_m \equiv c_1 c_2 \cdots c_m$ under 
map $\pi_2$, then $a_1 a_2 \cdots a_m \equiv c_1 c_2 \cdots c_m$
under map $\pi_2 \pi_1$.

\noindent $\it Proof$. 
From the assumption and Lemma 5,  $a_i \equiv b_i$ under the 
map $\pi_1$, for $i = 1, 2, \cdots, m$, and $b_i \equiv c_i$ under the
map $\pi_2$, for $i = 1, 2, \cdots, m$, thus $a_i \equiv c_i$ under 
the map $\pi_2\pi_1$, for $i = 1, 2, \cdots, m$. From Lemma 5 
again, $a_1 a_2 \cdots a_m \equiv c_1 c_2 \cdots c_m$ under map
$\pi_2\pi_1$.

Definition 4 and Definition 5 set up binary relations in the set of 
algorithms and pseudo-algorithms on
$\cup_{n=3}^{\infty}3SAT_{N}(n)$ 
respectively.  One of these binary relations can be extended to 
$\mathcal{TA}^m$ with $m \ge 1$. It is a equivalence relation 
by Corollary 3, Corollary 4 and Corollary 5. From Proposition 2,  
$\mathcal{TA}^1$ has only one equivalence class.  However, when 
$m \ge 2$, the following is true.

\noindent $\bf Proposition\mbox{ }3.$  For each $m \ge 2$,
$\mathcal{TA}^m$ has infinitely many equivalence classes.

\noindent $\it Proof$.  Consider the case $m = 2$ first.  Let $e_0^{-}$ 
be the negative generalized truth assignment, i.e., 
$e_0^{-} = e_1e_2 \cdots e_n \cdots$ 
where $e_n = \neg x_n^{\ast}$ for any integer $n \ge 1$.  Let 
$e_n^{-} = e_1e_2\cdots e_n \cdots$ be the generalized truth assignment,
where $e_k = \neg x_k^{\ast}$ for all $k \ge 1$ and $k \not= n$, and 
$e_n = x_n^{\ast}$.  For any integers $l \ge 1$, $k \ge 1$, if $l \not= k$, 
then $e_l^{-}e_0^{-} \not\equiv e_k^{-}e_0^{-}$.
Suppose that $e_l^{-}e_0^{-} \equiv e_k^{-}e_0^{-}$ under this condition,
from Lemma 5, $e_l^{-} \equiv e_k^{-}$
and $e_0^{-} \equiv e_0^{-}$ under the same map $\pi$.
However, from $e_0^{-} \equiv e_0^{-}$ and Lemma 4, $\pi$ is the
identical map, thus $e_l^{-} = e_k^{-}$. This contradicts to the condition
$l \not= k$. Since there are infinitely many integers greater than 1, the
proposition is true for case $m = 2$.

Now for the case $m > 2$, consider the following infinite sequence in 
$\mathcal{TA}^m$: 
\begin{eqnarray}
e_0^{-}\cdots e_0^{-}e_1^{-}, 
e_0^{-}\cdots e_0^{-}e_2^{-}, \cdots,
e_0^{-}\cdots e_0^{-}e_k^{-}, \cdots.
\label{eqn7.3}  
\end{eqnarray}
By the same argument as in the case $m = 2$, any two elements in 
(\ref{eqn7.3}) are not equivalent. So the proposition is true for this 
case as well.
  
Note that the proofs of Lemma 5, Corollary 3, Corollary 4 and 
Corollary 5 do not use Lemma 4. The argument in Lemma 5 can be 
generalized to the compositions of an algorithm and aggressive 
truth assignments, and the proofs of the following Corollary 6, 
Corollary 7 and Corollary 8 are similar to ones of Corollary 3, 
Corollary 4 and Corollary 5, respectively.  Thus, the concept of 
equivalence relation can be generalized from $\mathcal{TA}^\infty$ 
to $< f >$.

\noindent $\bf Lemma\mbox{ }6.$   For any given algorithm $\varphi$ 
on $\cup_{n=3}^{\infty}3SAT_{N}(n)$ and 
any $a_1, a_2, \cdots, a_m$, $b_1, b_2, \cdots, b_m$ $\in \mathcal{TA}^1$
with $m \ge 1$, $\varphi a_1 a_2 \cdots a_m \equiv \varphi b_1 b_2 \cdots b_m$ 
if and only if $\varphi \equiv \varphi$ and $a_i \equiv b_i$ under the 
same map $\pi$, for $i = 1, 2, \cdots, m$. 

\noindent $\bf Corollary\mbox{ }6.$  For any given algorithm $\varphi$ 
on $\cup_{n=3}^{\infty}3SAT_{N}(n)$ and
any $a_1, a_2, \cdots, a_m$ $\in \mathcal{TA}^1$ with $m \ge 1$,
$\varphi a_1 a_2 \cdots a_m \equiv \varphi a_1 a_2 \cdots a_m$ under 
the identical map.

\noindent $\bf Corollary\mbox{ }7.$  For any given algorithm $\varphi$ 
on $\cup_{n=3}^{\infty}3SAT_{N}(n)$, and
any 
$a_1, a_2, \cdots, a_m$ and $b_1$, $b_2, \cdots, b_m$, $\in \mathcal{TA}^1$ 
with $m \ge 1$, if
$\varphi a_1 a_2 \cdots a_m \equiv \varphi b_1 b_2 \cdots b_m$ under 
map $\pi$, then 
$\varphi b_1 b_2 \cdots b_m$ $\equiv \varphi a_1 a_2 \cdots a_m$ under 
map $\pi^{-1}$.

\noindent $\bf Corollary\mbox{ }8.$   For any given algorithm $\varphi$ 
on $\cup_{n=3}^{\infty}3SAT_{N}(n)$, 
and any $a_1, a_2, \cdots, a_m$, $b_1, b_2, \cdots$, $b_m$ and $
c_1, c_2, \cdots, c_m \in \mathcal{TA}^1$ with $m \ge 1$, if 
$\varphi a_1 a_2 \cdots a_m \equiv \varphi b_1 b_2 \cdots b_m$
under $\pi_1$ and 
$\varphi b_1 b_2 \cdots b_m \equiv \varphi c_1 c_2 \cdots c_m$
under $\pi_2$, then
$\varphi a_1 a_2 \cdots a_m \equiv \varphi c_1 c_2 \cdots c_m$
under $\pi_2\pi_1$.

\subsection*{8. Some Properties of Cauchy Sequences}

In this section, under the assumptions $\mathcal{A} \not= \emptyset$ and 
$\mathcal{A}a \subset \mathcal{A}$  for any $a \in {\mathcal{TA}}^1$, 
we discuss some properties of Cauchy sequences.

\noindent $\bf Definition\mbox{ }6.$ 
A Cauchy sequence $\{ f_n \}$ in
$< f >^2$ is called regular if it satisfies the following conditions:\\
(1) $f_n = f a_n a_0$  for $n = 1, 2, \cdots$; \\
(2) $a_0$ is an arbitrarily given aggressive truth assignment; \\
(3) $a_n$ and $a_{n+1}$ are identical on the first $n$ atomic truth assignments $e_1, e_2, \cdots, e_{n}$;\\ 
(4) $a_n$ and $a_0$ are identical on atomic truth assignments $e_{n+2}, e_{n+3}, \cdots$.

\noindent $\bf Remark\mbox{ }4.$  
From (3) and (4), $a_n$ and $a_{n+1}$ are identical on all atomic truth 
assignments but $e_{n+1}$ and $e_{n+2}$. Usually, the atomic truth 
assignments $e_{n+1}$ of $a_n$ and $e_{n+1}^0$ of $a_0$ are the 
same; however, in some cases, $e_{n+1}$ can be $\neg e_{n+1}^0$.  
The distance between $a_n$ and its adjustment $a_n^{\prime}$ is 
$d(a_n, a_n^{\prime}) = d(e_{n+1}^0, \neg e_{n+1}^0) = \frac{n+1}{2^{2n + 3}}$,
and $\sum_{n=1}^{\infty}\frac{n+1}{2^{2n + 3}}$ is convergent.
Thus a regular Cauchy sequence $\{ f a_n a_0 \}$ under any such 
adjustment is still a regular Cauchy sequence.

\noindent $\bf Example\mbox{ }4.$  Let $e_n = x_n^{\ast}$ for 
$n = 1, 2, \cdots$, $a_0 = \neg e_1 \neg e_2 \cdots$,  
the negative aggressive truth assignment, 
$a_n = e_1 \cdots e_{n+1} \neg e_{n+2} \neg e_{n+3} \cdots$,  
and $f_n = f a_n a_0$ for $n = 1, 2, \cdots$, then $\{ f_n \}$ is
a regular Cauchy sequence.  In this example, the $(n+1)th$ atomic 
truth assignments of $a_n$ and $a_0$ are different.

The following lemma claims that for any regular Cauchy sequence 
$\{ f_n \} \subset < f >^2$, there is an algorithm $g$ which takes 
more steps than each element of $\{ f_n \}$ does to determine if 
an arbitrarily given expression $\eta$ is satisfiable.

\noindent $\bf Lemma \mbox{ }7$.  
Let $\{ f_n \}$ in $< f >^2$ be a regular Cauchy sequence in Definition 6
and let $\tilde{a_0}$ be a less efficient aggressive truth assignment whose 
generalized truth assignment is identical with $a_0$.  For any 
$\eta \in \cup_{n=3}^{\infty}3SAT_{N}(n)$,
to determine if $\eta$ is satisfiable,  $f\tilde{a_0}\tilde{a_0}$ takes 
more steps than $f a_ka_0$ does, for $k = 1, 2, \cdots$.

\noindent $\it Proof$. 
From Lemma 1, $\tilde{a_0}$ takes more steps to evaluate $\eta$ 
than $a_0$ does. If $\tilde{a_0}$ returns {\it true},then $a_0$ 
returns {\it true} as well, both algorithms $f\tilde{a_0}\tilde{a_0}$ 
and $f a_ka_0$ complete, and $f\tilde{a_0}\tilde{a_0}$ takes 
more steps than $f a_ka_0$ does.  If  $a_0$ returns {\it false},  
then $\tilde{a_0}$ returns {\it false} as well.  In this case, using 
Lemma 1 again, $\tilde{a_0}$ takes more steps to evaluate $\eta$ 
than $a_k$ does, for $k = 1, 2, \cdots$, so to determine if $\eta$ 
is satisfiable $f \tilde{a_0}$ takes more steps than $f a_k$ does for 
$k = 1, 2, \cdots$ from (\ref{eqn6.3}).  Thus, the Lemma 7 follows.

\noindent $\bf Definition\mbox{ }7.$  
Let $\{ f_n \}, \{ f^{\prime}_n\} \in \mathcal{CS}$, 
$\{ f_n \}$ and $\{ f^{\prime}_n \}$ are equivalent if 
$f a_n a_0 \equiv f a^{\prime}_n a^{\prime}_0$ under 
bijective and ordered map $\pi_n$ for $n = 1, 2, \cdots$.

If $\{ f_n \}$ and $\{ f^{\prime}_n \}$ are equivalent, we write 
$\{ f_n \} \equiv \{ f^{\prime}_n \}$.  All $\pi_n$ in Definition 7 
are the same by Lemma 6 and  Lemma 4. Let $\mathcal{CS}$ be the
set of all regular Cauchy sequences in $< f >^2$.  Using the 
equivalent property of elements in $\mathcal{CS}$, we can set up a 
binary relation $\mathcal{R}$ in $\mathcal{CS}$ and prove that it is an 
equivalence relation by Corollary 6, Corollary 7 and Corollary 8.
In the following discussion, we can use Definition 7 in this way.   
Let $\mathcal{E_{CS}}$ be the set of all equivalence classes in 
$\mathcal{CS}$, i.e., 
$\mathcal{E_{CS}} = \mathcal{CS} / \mathcal{R}$. Let 
$\widetilde{\{ f_n \}}, \widetilde{\{ f^{\prime}_n\}} \in \mathcal{E_{CS}}$,  
$\{ f_n \} \in \widetilde{\{ f_n \}}$ and 
$\{ f^{\prime}_n \} \in \widetilde{\{ f^{\prime}_n\}}$,  if 
$\{ f_n \} = \{ f^{\prime}_n \}$, i.e., $\{ f_n \} \equiv \{ f^{\prime}_n \}$, 
then $f a_n a_0 \equiv f a^{\prime}_n a^{\prime}_0$ for 
$n = 1, 2, \cdots$, under the same map $\pi$.

\noindent $\bf Lemma \mbox{ }8$.  
Any element $\{ f_n \}$ in an equivalent class $\widetilde{\{ f_n \}}$ 
of $\mathcal{E_{CS}}$ represents a polynomial time algorithm on 
$\cup_{n=3}^{\infty}3SAT_{N}(n)$. 

\noindent $\it Proof$.
 From the assumption, any given regular Cauchy sequence $\{ f_n \}$ of 
$< f >^2$ is in $\mathcal{A}$ and converges to a point  $f_{\zeta}^{\ast}$.
It is not know if $f_{\zeta}^{\ast} \in \mathcal{A}$; however, it can be represented 
as an algorithm on $\cup_{n=3}^{\infty}3SAT_{N}(n)$. 
We may assume that $f_1 = f a_1 a_0, \cdots, f_n = f a_n a_0, \cdots$,
where $a_0$ is an aggressive truth assignment, $a_n$ and $a_0$ are identical 
after the $(n+1)th$ atomic truth assignment $e_{n+1}$ for $n = 1, 2, \cdots$. 
%%%%%%%%%%%%%%%%%%%%%%%%%%%%%%%%%%%%
For any instance 
$\eta \in \cup_{n=3}^{\infty}3SAT_{N}(n)$,
there exists an integer $n$ such that
$\eta \in 3SAT_{N}(n)$.
On the other hand, applying the polynomial time map $\phi_{map}$ 
to $\eta$, we get
$\eta^{\prime} = \phi_{map}(\eta) \in 3SAT_{N}(n)$.
Since 
$\phi_{map}$ is the identical map in $3SAT_{N}(n)$
for any integer $n \ge 3$, $\eta^{\prime} = \eta$. 
%%%%%%%%%%%%%%%%%%%%%%%%%%%%%%%%%%%%
For any given
$\eta \in \cup_{n=3}^{\infty}3SAT_{N}(n)$,
we can find the integer $n$ with
$\eta \in 3SAT_{N}(n)$ in polynomial time.
Thus, $f_{\zeta}^{\ast}$ can be represented as an algorithm on 
$\cup_{n=3}^{\infty}3SAT_{N}(n)$ 
as follows: for any $\eta \in 3SAT_{N}(n)$ and 
any $n \ge 3$,
\begin{eqnarray}
f_{\zeta}(\eta) = f_{n-2}(\eta) = f a_{n-2} a_0(\eta).
\label{eqn8.1}
\end{eqnarray}
Note that $a_0$ and $a_{n-2}$ are aggressive truth assignments. 
They can be described by finite information, thus $f_{\zeta}$  can be 
represented by finite information for each n.  By the assumption, $f$  
is a polynomial time algorithm, and $a_0$ and $a_n$ are polynomial 
time pseudo-algorithms, $f a_n a_0$ is a polynomial time algorithm.  
Let $\tilde{a_0}$ be the less efficient aggressive truth assignment 
whose generalized truth assignment is identical with $a_0$, 
$f \tilde{a_0} \tilde{a_0}$ is a polynomial time algorithm by the same 
reason,  and by Lemma 7,  it takes more steps than $f a_{n-2} a_0$ 
does for $n = 3, 4, \cdots$, to determine if an arbitrarily given 
expression $\eta$ is satisfiable. So $f_{\zeta}$ takes less steps 
than $f \tilde{a_0} \tilde{a_0}$ does to determine if an arbitrarily 
given expression $\eta$ is satisfiable.  Since $f$ and $f \tilde{a_0} \tilde{a_0}$ 
are both polynomial time algorithms,  $f_{\zeta}$ is a polynomial 
time algorithm as well.

\noindent $\bf Definition\mbox{ }8.$  The algorithm $f_{\zeta}$ 
constructed in Lemma 8 is called the representation of $\{ f_n \}$.

\noindent $\bf Definition\mbox{ }9.$ 
Let $\widetilde{\{ f_n \}}, \widetilde{\{ f^{\prime}_n \}} \in \mathcal{E_{CS}}$, 
$\{ f_n \} \in \widetilde{\{ f_n \}}$ and $\{ f^{\prime}_n \} \in \widetilde{\{ f^{\prime}_n \}}$, 
and $f_{\zeta}$ and $f^{\prime}_{\zeta}$ be their representations 
respectively,  $f_{\zeta}$ and $f^{\prime}_{\zeta}$ are equivalent if 
$f a_n a_0 \equiv f a^{\prime}_n a^{\prime}_0$ under some bijective 
and ordered map $\pi_n$ on $3SAT_{N}(n+ 2)$ for $n = 1, 2, \cdots$. 

\noindent $\bf Corollary\mbox{ }9.$  The representations of different 
elements in one equivalent class are equivalent.

\noindent $\it Proof$.  It follows directly from Lemma 8 and Definition 9.

\noindent $\bf Lemma \mbox{ }9$. 
Let $\widetilde{\{ f_n \}}, \widetilde{\{ f^{\prime}_n \}} \in \mathcal{E_{CS}}$, 
$\{ f_n \} \in \widetilde{\{ f_n \}}$ and $\{ f^{\prime}_n \} \in \widetilde{\{ f^{\prime}_n \}}$, 
and $f_{\zeta}$ and $f^{\prime}_{\zeta}$ be their representations respectively.  
If $\{ f_n \} \not\equiv \{ f^{\prime}_n \}$, then $f_{\zeta} \not\equiv f^{\prime}_{\zeta}$.

\noindent $\it Proof$. We use the proof by contradiction. 
Suppose that $f_{\zeta} \equiv f^{\prime}_{\zeta}$. By Definition 9, 
$f \equiv f$ and $a_0 \equiv a_0^{\prime}$, and $a_n \equiv a_n^{\prime}$ 
under the bijective and ordered map $\pi_n$ on $3SAT_{N}(n+ 2)$ 
for $n = 1, 2, \cdots$.  By Definition 6, $a_n$ and $a_0$, and $a_n^{\prime}$ 
and $a_0^{\prime}$ have the same extensions after the (n+1)$th$ atomic 
truth assignments respectively, so $a_n \equiv a_n^{\prime}$  and 
$a_0 \equiv a_0^{\prime}$ on $3SAT_{N}(n+ 2)$ can be extended to
$\cup_{n=3}^{\infty}3SAT_{N}(n)$, i.e., $a_n \equiv a_n^{\prime}$ 
and $a_0 \equiv a_0^{\prime}$  under the extended bijective and ordered 
map $\pi_n$.  This is true for $n = 1, 2, \cdots$.  On the other hand, 
all $\pi_n$ are the same by Lemma 6 and Lemma 4.
Let $\pi$ be this map, then $f \equiv f$ under map $\pi$.
Thus, $\{ f_n \} \equiv \{ f^{\prime}_n \}$. This
contradicts to the assumption $\{ f_n \} \not\equiv \{ f^{\prime}_n \}$,
and hence $f_{\zeta} \not\equiv f^{\prime}_{\zeta}$.

\noindent $\bf Corollary\mbox{ }10.$
Let $\widetilde{\{ f_n \}}, \widetilde{\{ f^{\prime}_n \}} \in \mathcal{E_{CS}}$, 
$\{ f_n \} \in \widetilde{\{ f_n \}}$ and $\{ f^{\prime}_n \} \in \widetilde{\{ f^{\prime}_n \}}$, 
and $f_{\zeta}$ and $f^{\prime}_{\zeta}$ be their representations respectively.  
If $\{ f_n \} \not\equiv \{ f^{\prime}_n \}$, then $f_{\zeta} \not= f^{\prime}_{\zeta}$.

\noindent $\it Proof$. 
First of all, we have $f_{\zeta} \not\equiv f^{\prime}_{\zeta}$ by Lemma 9.  
From Definition 9, there exists an integer $n$, such that 
$fa_na_0 \not\equiv fa^{\prime}_na^{\prime}_0$ under any bijective and 
ordered map $\pi_n$ on $3SAT_{N}(n+ 2)$. This implies that 
$a_n \not= a^{\prime}_n$ or $a_0 \not= a^{\prime}_0$ on the first 
$n + 2$ atomic truth assignments. If $a_0 \not= a^{\prime}_0$ on the 
first $n + 2$ atomic truth assignments, then $a_0 \not= a^{\prime}_0$ 
as an aggressive truth assignment. From Lemma 3, for any $m \ge n$, 
there exists an $\eta_m \in 3SAT_N(m)$ such that 
$a_0(\eta_m) = true$ and $a^{\prime}_0(\eta_m) = false$. Thus, 
$f_{\zeta} \not= f^{\prime}_{\zeta}$ under the sense of algorithms.  Now 
we may assume that $a_0 = a^{\prime}_0$ as aggressive truth assignments 
and $a_n \not= a^{\prime}_n$ on the first $n + 2$ atomic truth assignments. 
We may further assume that $a_n \not= a_0$.   From Definition 6, for any 
$m \ge n+ 2$, $a_m \not= a^{\prime}_m$ on the first $m + 2$ atomic 
truth assignments.  As discussed above, using Lemma 3, we can show that 
$f_{\zeta} \not= f^{\prime}_{\zeta}$ as algorithms.

\subsection*{9. Proof of the Proposition 1}
In this section, we prove Proposition 1 which claims that
$\mathcal{AT}^{1}$ is compatible with the $\mathcal P$ versus 
$\mathcal {NP}$ problem. From the definition of aggressive truth
assignment and equation (\ref{eqn6.3}), each element of 
$\mathcal{TA}^{1}$ satisfies conditions (1), (2) and (3) of 
Definition 1. We just prove here that $\mathcal{TA}^{1}$ 
satisfies condition (4) of Definition 1 as well.

To begin with, we use some elements of ${\mathcal{TA}}^1$ to construct
an exponential time algorithm on $3SAT_{N}(n)$ for each $n \ge 3$. Without 
lost of generality, we may assume that all elements of ${\mathcal{TA}}^1$
used in this proof have the negative extension.  
Let
\begin{eqnarray}
(e_1 e_2 ... e_n)^{2^n} = (e_1^{1} e_2^{1}...e_n^{1})
(e_1^{2} e_2^{2}...e_n^{2})\cdots (e_1^{2^n} e_2^{2^n}...e_n^{2^n}), \nonumber
\end{eqnarray}
where $e_1^{i}e_2^{i} \cdots e_n^{i}$ $\not= e_1^{j}e_2^{j} \cdots e_n^{j}$,
under the sense of regular truth assignment, if $i \not= j$.
Let
$e_{n+1}^{1} = \cdots = e_{n+1}^{2^n} = \neg x_{n+1}^{\ast}$ and
$e_{n+1}^{2^n+1} =  \cdots = e_{n+1}^{2^{n+1}} = x_{n+1}^{\ast}$, then
\begin{eqnarray}
(e_1 \cdots e_n)^{2^n}
&=& (e_1^{1} \cdots e_n^{1})
(e_1^{2} \cdots e_n^{2})
\cdots (e_1^{2^n} \cdots e_n^{2^n}) \nonumber \\
&=& (e_1^{1} \cdots e_n^{1} \neg x_{n+1}^{\ast})
(e_1^{2} \cdots e_n^{2} \neg x_{n+1}^{\ast})
\cdots (e_1^{2^n} \cdots e_n^{2^n} \neg x_{n+1}^{\ast}) \nonumber \\
&=& (e_1^{1} \cdots e_{n+1}^{1})
(e_1^{2} \cdots e_{n+1}^{2}) \cdots 
(e_1^{2^n} \cdots e_{n+1}^{2^n}). \nonumber
\end{eqnarray}
Let $(e_1 ... e_n x_{n+1}^{\ast})^{2^n} 
= (e_1^{2^n + 1} \cdots e_n^{2^n + 1} x_{n+1}^{\ast})
(e_1^{2^n + 2} \cdots e_n^{2^n + 2} x_{n+1}^{\ast})
\cdots (e_1^{2^{n+1}} \cdots e_n^{2^{n+1}} x_{n+1}^{\ast})$, then
\begin{eqnarray}
&&(e_1 ... e_n x_{n+1}^{\ast})^{2^n} 
= (e_1^{2^n + 1} \cdots e_{n+1}^{2^n + 1})
(e_1^{2^n + 2} \cdots e_{n+1}^{2^n + 2}) \cdots 
(e_1^{2^{n+1}} \cdots e_{n+1}^{2^{n+1}}), \nonumber \\
&&(e_1 e_2 ... e_n)^{2^n}(e_1 ... e_n x_{n+1}^{\ast})^{2^n}
= (e_1 e_2 ... e_{n+1})^{2^{n+1}}.
\label{eqn9.1}
\end{eqnarray}
We can construct a sequence in $< f >$ by (\ref{eqn9.1}) inductively: let 
$(e_1 e_2e_3)^{2^3}$ =
$(\neg x_1^{\ast} \neg x_2^{\ast} \neg x_3^{\ast})$
$(x_1^{\ast}\neg x_2^{\ast}\neg x_3^{\ast})$
$(\neg x_1^{\ast} x_2^{\ast} \neg x_3^{\ast})$
$(x_1^{\ast} x_2^{\ast} \neg x_3^{\ast})$
$(\neg x_1^{\ast} \neg x_2^{\ast} x_3^{\ast})$
$(x_1^{\ast}\neg x_2^{\ast} x_3^{\ast})$
$(\neg x_1^{\ast} x_2^{\ast} x_3^{\ast})$
$(x_1^{\ast} x_2^{\ast} x_3^{\ast})$,
\begin{eqnarray}
f_1 &=& f (e_1 e_2e_3)^{2^3}, \nonumber \\
f_2 &=& f_1 (e_1 e_2 e_3 x_4^{\ast})^{2^3} = f (e_1 ... e_4)^{2^4}, \cdots , \nonumber \\
f_k &=& f_{k-1} (e_1 \cdots e_{k+1} x_{k+2}^{\ast})^{2^{k+1}} = f (e_1 ... e_{k+2})^{2^{k+2}}, \cdots . \nonumber
\end{eqnarray}
Clearly, $f_k \in < f >^{2^{k+2}}$ for any integer $k \ge 1$. 
Since series $\sum_{k=1}^{\infty} \frac{1}{k^2}$ is convergent,
for any real number $\epsilon > 0$, there exists a
positive integer $N$, such that 
$\sum_{k=N}^{\infty} \frac{1}{k^2} < \epsilon$.
Thus, for all natural numbers $m, n > N$, 
$d(f_m, f_n) = d(f(e_1 \cdots e_{m+2})^{2^{m+2}}, f(e_1 \cdots e_{n+2})^{2^{n+2}} )< \sum_{k=N}^{\infty} \frac{1}{k^2} < \epsilon$, 
so $\{f_n\}$ is a Cauchy sequence in $< f >$.
Suppose that $\{f_n\}$ converges to a point $f_\xi$.

For any instance 
$\eta \in \cup_{n=3}^{\infty}3SAT_{N}(n)$,
there exists an integer $n$ such that $\eta \in 3SAT_{N}(n)$.
On the other hand, applying the polynomial time map $\phi_{map}$ 
to $\eta$, we get
$\eta^{\prime} = \phi_{map}(\eta) \in 3SAT_{N}(n)$.
Since 
$\phi_{map}$ is the identical map in $3SAT_{N}(n)$
for any integer $n \ge 3$, $\eta^{\prime} = \eta$. Thus, for any 
$\eta \in \cup_{n=3}^{\infty}3SAT_{N}(n)$,
we can find the integer $n$ with $\eta \in 3SAT_{N}(n)$ 
in polynomial time.  From the construction, 
\begin{eqnarray}
f_\xi (\eta) = f_{n-2}(\eta) = (e_1e_2 \cdots e_n)^{2^n}(\eta),
\label{eqn9.2}
\end{eqnarray}
for any $\eta \in 3SAT_{N}(n)$ and any integer $n \ge 3$. 
From above discussion, $(e_1e_2 \cdots e_n)^{2^n}$ can be described
by finite information for each $n$. Thus, $f_\xi$ can be represented by 
finite information for each $n \ge 3$. 

From the construction,  $(e_1e_2 \cdots e_n)^{2^n}$ is an 
algorithm on $3SAT_{N}(n)$ for each $n \ge 3$ and its last
aggressive truth assignment is negative.  Let 
$a^{\ast} \in {\mathcal{TA}}^1$ be the negative aggressive 
truth assignment. From Lemma 3, there exists an 
$\eta_n \in 3SAT_{N}(n)$ for each $n \ge 3$, such that 
$a^{\ast}(\eta_n) = true$ and $a(\eta_n) = false$ for all 
other $a \in {\mathcal{TA}}^1$ which is not equal to 
$a^{\ast}$ in the first $n$ atomic truth assignments. Now 
$(e_1e_2 \cdots e_n)^{2^n}(\eta_n) = true$; however, 
the algorithm must use all its $2^n$ aggressive truth assignments to 
evaluate $\eta_n$. From the proof of Lemma 3, 
$\eta_n$ has $4n$ clauses. Therefore, $(e_1e_2 \cdots e_n)^{2^n}$ 
takes exponential time to evaluate $\eta_n$ in the length of 
$\eta_n$ for each $n \ge 3$, so the equation (\ref{eqn9.2}) is an 
exponential time algorithm on $\cup_{n=3}^{\infty}3SAT_{N}(n)$.
From Lemma 3, we cannot reduce the time complexity of 
$(e_1e_2 \cdots e_n)^{2^n}$ by taking any aggressive 
truth assignments out from it.  The proof of Proposition 1 completes.

By the same argument, the time complexity of any algorithm 
generated by a similar construction using elements of ${\mathcal{TA}}^1$ 
is always at least exponential.

\subsection*{10. Proof of the Main Result}

In this section, we prove that $\mathcal A$ is empty using proof by 
contradiction. Suppose that $\mathcal A$ is not empty. Let $f_\xi$ be 
a polynomial time algorithm on 
$\cup_{n=3}^{\infty}3SAT_{N}(n)$ and
$a$ be an aggressive truth assignment, then $f_\xi a$ is an algorithm 
on $\cup_{n=3}^{\infty}3SAT_{N}(n)$. 
From the previous sections, $f_\xi a$ is a polynomial time algorithm 
as well.  However, it is not clear if $f_\psi a \in \mathcal A$ for any 
$f_\psi \in \mathcal A$ and any aggressive truth assignment $a$. 
Without loss of generality, we may take $f$ as this $f_\psi$ in the 
previous sections.  There are two cases:

\noindent (1) for any aggressive truth assignment $a$, $\mathcal A a \subseteq \mathcal A$;

\noindent (2) there is an aggressive truth assignment $a_{\ast}$ such that
$\mathcal A a_{\ast}  \not\subset \mathcal A$.

We prove that neither case (1) nor case (2) is true in this section.

\noindent $\it Case$ $\it (1)$.
Under the assumptions, we want to prove in this case that there exist 
uncountably many polynomial time algorithms in $\mathcal{A}$.
In the following discussion, we just consider the equivalence classes of 
$\mathcal{CS}$, i.e., the elements in $\mathcal{E_{CS}}$.  Since each 
aggressive truth assignment is defined by finite information, 
${\mathcal{TA}}^1$ is countable; however, there is only one element in 
${\mathcal{TA}}^1$ under the equivalence relation. In general, 
$< f >^1$ is countable under the equivalence relation. We want to 
prove that $\mathcal{E_{CS}}$ is uncountable. Suppose that 
$\mathcal{E_{CS}}$ is countable, all elements of $\mathcal{E_{CS}}$ 
can be listed as $\widetilde{\{f_n^1\}}$, $\widetilde{\{f_n^2\}}$, 
$\cdots, \widetilde{\{f_n^k\}}, \cdots$.  We can chooes one element 
$\{ f_n^k\}$ in each equivalent class $\widetilde{\{f_n^k\}}$ as follows:
\begin{eqnarray}
\{f_n^1\} &=& fa_1^1a_0^1, fa_2^1a_0^1, \cdots, fa_k^1a_0^1, \cdots, \nonumber \\ 
\{f_n^2\} &=& fa_1^2a_0^2, fa_2^2a_0^2, \cdots, fa_k^2a_0^2, \cdots, \nonumber \\ 
&& \cdots, \label{eqn10.1} \\
\{f_n^k\} &=& fa_1^ka_0^k, fa_2^ka_0^k, \cdots, fa_k^ka_0^k, \cdots, \nonumber \\ 
&&\cdots. \nonumber
\end{eqnarray}
Suppose that 
\begin{eqnarray}
a_1^1 &=& e_1^1 e_2^1 \cdots e_k^1 \cdots, \nonumber \\
a_2^2 &=& e_1^2 e_2^2 \cdots e_k^2 \cdots, \nonumber \\
&& \cdots, \label{eqn10.2} \\
a_k^k &=& e_1^k e_2^k \cdots e_k^k \cdots, \nonumber \\
&& \cdots. \nonumber
\end{eqnarray}
Define
\begin{eqnarray}
a_1 &=& \neg e_1^1 e_2 \neg x_3^{\ast} \cdots \neg x_k^{\ast} \neg x_{k+1}^{\ast} \cdots, \nonumber \\
a_2 &=& \neg e_1^1 \neg e_2^2 e_3 \neg x_4^{\ast} \cdots \neg x_k^{\ast} \neg x_{k+1}^{\ast} \cdots, \nonumber \\
&& \cdots, \label{eqn10.3} \\
a_k &=& \neg e_1^1 \neg e_2^2 \cdots \neg e_k^k e_{k+1} \neg x_{k+2}^{\ast} \cdots, \nonumber \\
&& \cdots, \nonumber
\end{eqnarray}
where $e_m = x_m^{\ast} \mbox{ or } \neg x_m^{\ast}$ for $m \ge 2$, and
\begin{eqnarray}
\{ f_n \} = f a_1a_0, f a_2a_0, \cdots, fa_na_0, \cdots, \label{eqn10.4}
\end{eqnarray}
where $a_0$ is the aggressive truth assignment with the negative generalized 
truth assignment.  Note that $\{ f_n \} \subset < f >^2$ is a regular Cauchy 
sequence from Lemma 10.  We can adjust the aggressive truth assignments in 
(\ref{eqn10.3}) if necessary.  For example, if $a_0 \equiv a_0^1$ and 
$a_1 \equiv a^1_1$ under the same map $\pi_1$, we can adjust 
$a_1 =  \neg e_1^1 \neg e_2 \neg x_3^{\ast} \cdots \neg x_n^{\ast} \cdots$.
From Lemma 4, $a_0 \equiv a_0^1$ and $a_1 \equiv a^1_1$,
but under different maps.  Similarly, if $a_0 \equiv a_0^2$ and $a_2 \equiv a^2_2$
under the same map $\pi_2$, we can adjust 
$a_2 = \neg e_1^1 \neg e_2^2 \neg e_3 \neg x_4^{\ast} \cdots \neg x_n^{\ast} \cdots$
such that $a_0 \equiv a_0^2$ and $a_2 \equiv a^2_2$ under different maps.
This modification has no impact on the previous one, i.e., the claim 
$a_0 \equiv a_0^1$ and $a_1 \equiv a^1_1$ under different maps is still 
true. Sequence $\{ a_n \}$ is still a Cauchy sequence by Remark 4
and satisfies conditions (3) and (4) of Definition 6. Sequence $\{ f_n \}$ is still a 
regular Cauchy sequence. From the assumption, $\{ f_n \} \subset \mathcal{A}$. 
Thus, $\{ f_n \} \in \mathcal{CS}$; however, from the construction, 
$fa_1a_0 \not= fa_1^1a_0^1, fa_2a_0 \not= fa_2^2a_0^2, \cdots, 
fa_k a_0 \not= fa_k^k a_0^k, \cdots$, under the equivalence relation, i.e., 
$\{ f_n \}$ is not in the list (\ref{eqn10.1}).  This is a contradiction, which 
implies that $\mathcal{E_{CS}}$ is uncountable. Any element of $\mathcal{E_{CS}}$ 
has a family of representations which are polynomial time algorithms on 
$\cup_{n=3}^{\infty}3SAT_{N}(n)$ and 
equivalent each other by Lemma 8 and Corollary 9.  The representations 
of elements in different equivalent classes in $\mathcal{E_{CS}}$ are 
not equivalent by Lemma 9, and they are different as algorithms
by Corollary 10.  Therefore, there exist uncountably many 
algorithms in $\mathcal{A}$. This is absurd, since there 
are only countably many algorithms (see e.g. \cite{Hein1}).  So case (1) 
is not true. 

\noindent $\it Case$ $\it (2)$.
From the assumption, there exist an $f_\lambda \in \mathcal{A}$ and an 
$a_{\ast} \in {\mathcal{TA}}^1$, such that $f_\lambda a_{\ast} \not\in \mathcal{A}$.
Since $a_{\ast}$ is a polynomial time pseudo-algorithm and $f_\lambda$ is a 
polynomial time algorithm on 
$\cup_{n=3}^{\infty}3SAT_{N}(n)$, 
$f_\lambda a_{\ast}$ is a polynomial time algorithm on
$\cup_{n=3}^{\infty}3SAT_{N}(n)$, 
i.e., $f_\lambda a_{\ast} \in \mathcal{A}$.  This is a contradiction. So Case 2 
is not true, either.

\noindent $\bf Remark\mbox{ }5.$  It is not necessary to prove that 
the representation of a regular Cauchy sequence is a polynomial time 
algorithm in Lemma 8.  As long as any representation is an algorithm, 
and Lemma 9 is true, the argument in Case (1) is still valid, i.e.,
there exist uncountably many algorithms on
$\cup_{n=3}^{\infty}3SAT_{N}(n)$.
This is a contradiction too. However, being able to prove that any 
representation is a polynomial time algorithm reveals more properties 
of  $\mathcal{A}$ and helps us to understand the polynomial time
algorithms better.

\noindent $\bf Lemma \mbox{ }10$. The sequence $\{ f_n \}$ defined in 
(\ref{eqn10.4}) is a regular Cauchy sequence.  

\noindent $\it Proof$. 
Since $a_0$ in the algorithm sequence (\ref{eqn10.4}) is the aggressive 
truth assignment with the negative generalized truth assignment,  
$\{ f_n \}$ satisfies the condition (1) of Definition 6.  The constructions 
(\ref{eqn10.2}) and (\ref{eqn10.3}) show that $\{ f_n \}$ satisfies 
conditions (2), (3) and (4) of Definition 6.  In order to complete the 
proof, we must prove that $\{ f_n \}$ is a Cauchy sequence.  For any 
given $\varepsilon > 0$, since $\sum_{n = 1}^{\infty}\frac{n}{2^{2n+1}}$
is convergent, there exists a positive integer $N$, such that 
$\sum_{n = N}^{\infty}\frac{n}{2^{2n+1}} < \varepsilon$.
Thus, if $k > N$ and $l \ge 0$, from (\ref{eqn10.3}), 
\begin{eqnarray}
d(a_k, a_{k + l}) = d(e_{k+1}, \neg e_{k+1}^{k+1}) +  
                                \sum_{n = k+2}^{k + l}d(\neg x_n^{\ast}, \neg e_n^n) +
                                d(\neg x_{k+ l+1}^{\ast}, e_{k+ l+1})
                                \le \sum_{n = k+1}^{k + l + 1}\frac{n}{2^{2n+1}} < \varepsilon. \nonumber
\end{eqnarray}
Therefore, if $n, m > N$, 
$d(f a_n a_0, f a_m a_0) = d(a_n a_0, a_m a_0) = d(a_n, a_m) < \varepsilon$ 
from (\ref{eqn5.5}) and (\ref{eqn6.5}), i.e., $\{ f_n \}$ is a Cauchy sequence.

\noindent $\bf Lemma \mbox{ }11$. $\mathcal A = \emptyset$.

\noindent $\it Proof$. It follows directly from the above discussion.

We know that $\cup_{n=3}^{\infty}3SAT_{N}(n)$ is $\mathcal {NP}$-complete.
Lemma 11 claims that there does not exist any polynomial time algorithm on
$\cup_{n=3}^{\infty}3SAT_{N}(n)$,
so $\cup_{n=3}^{\infty}3SAT_{N}(n) \not\in \mathcal P$.
We have the following result:

\noindent $\bf Theorem \mbox{ }3$. $\mathcal {P} \ne \mathcal {NP}$. 

We can apply the new argument to $2SAT$; however, we cannot get 
any contradiction, so we cannot change the status of $2SAT$. Since 
$2SAT \in \mathcal{P}$, if we classify $2SAT$, we obtain $2SAT$ 
itself or some classes in $2SAT$ which are also in $\mathcal{P}$. 
For any such classification, we can define an aggressive truth
assignment $e_1e_2\cdots e_m$ as follows:
for any $\eta \in 2SAT(n)$,
(1) it evaluates $\eta$ as a generalized truth assignment;
(2) it checks that if $\eta$ is satisfiable using
    Aspvall, Plass and Tarjan's algorithm \cite{Aspvall1}.
Now for any instance $\eta \in 2SAT$, the aggressive 
truth assignment $e_1e_2 \cdots e_m$ works in this way: 
(1) it evaluates $\eta$ as a truth assignment, 
  if $e_1e_2 \cdots e_m(\eta) = true$,
  it returns a true value, otherwise it goes to next step; 
(2) it checks if $\eta$ is satisfiable using Aspvall, Plass and 
  Tarjan's algorithm \cite{Aspvall1}. If $\eta$ is satisfiable, 
  it returns a true value, otherwise it returns a false value. 
So $e_1e_2 \cdots e_m(\eta) = true$ if and only if $\eta$ is satisfiable.  
Since Aspvall, Plass and Tarjan's algorithm is linear time algorithm on 
$2SAT$, all aggressive truth assignments are linear time algorithms on 
$2SAT$ as well. Unlike $3SAT_{N}$, we cannot introduce the compatible set concept in 
this case.  The pseudo-algorithm concept is suitable for $3SAT_{N}$, 
but makes no sense for $2SAT$.

\subsection*{Acknowledgements}
It is my pleasure to thank Dr. Gary R. Jensen for his  
encouragement. I am also very grateful to Dr. Yining Wang, who
has contributed many ideas and comments. I greatly appreciate 
my wife, Ms. Yongmei Feng, and my children for their 
understanding, patience, support and love.

\end{document}